\documentclass[12pt]{article}
\usepackage{amsmath}
\usepackage{amsthm}
\usepackage{graphicx}
\usepackage{enumerate}
\usepackage{natbib}
\usepackage{algorithm}
\usepackage{algorithmic}
\usepackage{setspace}
\usepackage{amssymb}
\usepackage{url} 
\usepackage{multirow}
\usepackage{booktabs}
\usepackage{array}
\usepackage[justification=raggedright,singlelinecheck=false]{subcaption}
\usepackage{colortbl}
\usepackage{bbding}
\usepackage{lscape}
\usepackage{import}
\usepackage{booktabs}
\usepackage{longtable}
\usepackage[normalem]{ulem}
\usepackage{makecell}
\usepackage{tabularx}
\usepackage{adjustbox}
\usepackage{comment}

\newtheoremstyle{italichead}  
  {\topsep}                   
  {\topsep}                   
  {\normalfont}               
  {}                          
  {\itshape}                  
  {.}                         
  {.5em}                      
  {\thmname{\itshape #1} \ \thmnumber{\itshape #2}\thmnote{ (#3)}} 

\theoremstyle{italichead}

\newtheorem{example}{Example}[section]

\begin{document}

\def\spacingset#1{\renewcommand{\baselinestretch}%
{#1}\small\normalsize} \spacingset{1}


\title{\bf Dominating Hyperplane Regularization for Variable Selection in Multivariate Count Regression}

\author{
  Alysha Cooper\textsuperscript{1}\thanks{This work was supported by the Oil Sands Monitoring Program via the Government of Alberta under Grant 21GRRSD14-02; Discovery Grant from the Natural Sciences and Engineering Research Council of Canada under Grants 400095, 2019–04204 (ZF) and 400808, 2021-02856 (AA).}, 
  Zeny Feng\textsuperscript{1}, 
  Ayesha Ali\textsuperscript{1}, 
  Tim Arciszewski\textsuperscript{2}, 
  Lorna Deeth\textsuperscript{1}
}

\date{
\textsuperscript{1}Department of Mathematics and Statistics, University of Guelph\\
\textsuperscript{2}Alberta Environment and Protected Areas
}

\maketitle

\begin{abstract}
Identifying relevant factors that influence the multinomial counts in compositional data is difficult in high dimensional settings due to the complex associations and overdispersion. Multivariate count models such as the Dirichlet-multinomial (DM), negative multinomial, and generalized DM accommodate overdispersion but are difficult to optimize due to their non-concave likelihood functions. Further, for the class of regression models that associate covariates to the multivariate count outcomes, variable selection becomes necessary as the number of potentially relevant factors becomes large. The sparse group lasso (SGL) is a natural choice for regularizing these models. Motivated by understanding the associations between water quality and benthic macroinvertebrate compositions in Canada’s Athabasca oil sands region, we develop dominating hyperplane regularization (DHR), a novel method for optimizing regularized regression models with the SGL penalty. Under the majorization-minimization framework, we show that applying DHR to a SGL penalty gives rise to a surrogate function that can be expressed as a weighted ridge penalty. Consequently, we prove that for multivariate count regression models with the SGL penalty, the optimization leads to an iteratively reweighted Poisson ridge regression. We demonstrate stable optimization and high performance of our algorithm through simulation and real world application to benthic macroinvertebrate compositions.
\end{abstract}

\noindent%
\vfill

\section{Introduction}

\label{sec:intro}

Multivariate count data, measured by taxa counts at a specified taxonomic rank, are prevalent in many biological fields including microbiology, genetics, and ecology. In these biological fields, one may collect samples from different locations or subjects, classify organisms (such as benthic macroinvertebrates in ecology or gut bacteria in microbiology) at a given taxonomic rank, and then count the number of each taxon observed in the sample relative to the total count. In this study, we aim to identify important water quality variables associated with the composition of benthic macroinvertebrate living in the Athabasca oil sands region in Alberta, Canada. Benthic communities are sensitive to pollution \citep{kroncke2010influence} and therefore are used as indicators of the impacts of pollutants and stressors present in the Athabasca oil sands region. Unfortunately, identifying water quality factors that are associated with the abundance of each taxon is a complicated task. 

Conditional on the total number of organisms observed in a sample (i.e., total count), the natural distribution for the abundances of the taxa is the multinomial distribution. However, in practice, the multivariate counts often exhibit greater variability than what is expected under the multinomial distribution assumption. To account for this increased variability, the proportion parameters of the multinomial model can be treated as a vector of random variables following a Dirichlet distribution, giving rise to a Dirichlet-multinomial (DM) distribution \citep{mosimann1962compound}. When there are covariates that influence the compositional distribution of the counts, DM regression can be used to model the multivariate count composition. A major drawback to the DM regression model is its non-concave log-likelihood function, which makes finding a maximum likelihood estimate (MLE) solution difficult via traditional estimation methods.

When using DM regression to model the relationship between the compositional distribution of multivariate counts and covariates, there are two sources of dimensionality: 1) the number of potentially relevant covariates, denoted by $p$; and 2) the number of taxa, denoted by $D$. Consequently, there are $(p+1)\times D$ coefficients, including the intercept terms. The coefficient $\beta_{jd}$ quantifies the association between the $j^{th}$ covariate and the count of the $d^{th}$ taxon. Performing variable selection among the coefficients in the $(p+1)\times D$ coefficient matrix, denoted $\boldsymbol{\beta}$, becomes necessary as $p$ grows large. Regularization, a stable method for variable selection, involves the addition of a penalty term to the objective function during optimization to favour more parsimonious models.  In regularized regression problems, we seek to minimize the objective function $-\ell(\boldsymbol{\beta}) + \lambda J(\boldsymbol{\beta})$, where $\ell(\boldsymbol{\beta})$ is the log-likelihood function, $J(\boldsymbol{\beta})$ is a penalty function, and $\lambda$ is a tuning parameter used to determine the trade-off between model fit and model complexity. The choice of penalty function depends on the overall goal of the model.

Coefficient parameters in a DM regression model have a natural grouping structure. Let $\boldsymbol{\beta}_j$ denote the $D$-length vector from the $j^{th}$ row of the $(p+1)\times D$ matrix $\boldsymbol{\beta}$. The effects of a covariate across all outcomes can be organized in a group and the group lasso penalty can set all coefficients of a covariate to 0 across all taxa (i.e., $\boldsymbol{\beta}_{j} = \boldsymbol{0}$) rather than shrinking the individual coefficients for that covariate \citep{yuan2006model}. In addition, the group lasso penalty can be combined with the lasso penalty \citep{tibshirani1996regression} to form the \textit{sparse group lasso} (SGL) penalty \citep{simon2013sparse}, in which individual coefficients within the remaining groups can be shrunk to zero. Suppose we have $m$ groups of coefficients with $D_j$ coefficients for groups $j = 1, 2, \ldots, m$. We seek to minimize the objective function
\begin{eqnarray}
\label{eqn:sgl}
    f(\boldsymbol{\beta}) & = & -\ell(\boldsymbol{\beta}) + \lambda_1||\beta||_1+\lambda_2\sum_{j=1}^m\sqrt{D_j}||\beta_j||_2  \\
    & = & -\ell(\boldsymbol{\beta}) +  \alpha\lambda\sum_{j=1}^m \sum_{d=1}^{D_j}|\beta_{jd}| + (1-\alpha)\lambda\sum_{j=1}^m  \sqrt{D_j}   \sqrt{\sum_{d=1}^{D_j}\beta_{jd}^2} \nonumber,
\end{eqnarray}
where $\lambda_1$ and $\lambda_2$ are the tuning parameters and $\alpha = \frac{\lambda_1}{\lambda_1+\lambda_2}$ and $\lambda = \lambda_1 + \lambda_2$ are re-parameterizations of $\lambda_1$ and $\lambda_2$ such that $\alpha$ ranges between 0 and 1. Consequently, $\alpha$ determines the balance between group selection (group lasso) and within-group selection (sparse lasso) such that $\alpha=0$ is the group lasso and $\alpha=1$ is the sparse lasso. The resulting model enhances interpretability at both the group level and the within-group level. 

In the motivating benthic macroinvertebrate example, a regularized DM regression with the SGL penalty can be applied to identify important water quality variables predictive of benthic macroinvertebrate composition at a specified taxonomic rank. The Dirichlet prior can accommodate any observed multinomial overdispersion while the associations retained after applying the SGL can inform development of effective water conservation and management policies. However, SGL for DM regression cannot easily be conducted due to the multivariate outcome and the poorly behaved objective function (e.g., non-smooth, non-convex). Traditional optimization methods such as coordinate descent and gradient descent methods excel with smooth and strictly convex objective functions but falter with non-smooth or non-convex objective functions due to hindered convergence in regions with near-zero slope. 

The minorization-maximization and the majorization-minimization algorithms, both referred to as the MM algorithm, are promising alternative methods for optimizing complex objective functions \citep{zhang2017regression, wu2010mm} when the conventional descent/ascent-based methods are not satisfactory. The MM algorithm relies on iteratively optimizing a simple surrogate function that is tangential to and bounded by the objective function. This algorithm is stable, adaptable to parameter constraints, scalable to high dimensions, and is able to separate model parameters. For example, a surrogate function for the non-regularized DM regression at each iteration is the sum of iteratively reweighted Poisson regressions over $D$ taxa \citep{zhang2017regression}. This optimization algorithm is referred to as the iteratively re-weighted Poisson regression (IRPR).   

Motivated by the gap in existing algorithms for stable optimization of the regularized DM regression and inspired by the success of the MM algorithm in optimizing the non-regularized DM regression, we propose dominating hyperplane regularization (DHR). DHR constructs a majorizing surrogate function via the dominating hyperplane inequality for the SGL penalty shown in Eq. (\ref{eqn:sgl}). Combining the DHR surrogate for the penalty function with the IRPR surrogate for non-regularized DM regression, we develop a novel MM algorithm to facilitate the optimization of the regularized DM regression. Our DHR surrogate function for the penalty can be expressed as a weighted ridge $L_2$ penalty, and consequently, the surrogate function of the regularized DM regression with the SGL penalty can be expressed as the sum over $D$ iteratively reweighted Poisson ridge regressions. Unlike previous optimization methods for the SGL, this simple and elegant algorithm optimizes regularized likelihoods without requiring calculation of complex first- or second-order derivatives of the log-likelihood with respect to each of the regression parameters \citep{chen2013variable,zhang2017regression}. While our interest lies in regularized multivariate count regression, DHR presents a general framework for fitting regularized regressions of other distributional models with the SGL or other choices of penalty functions.   


\subsection{Relation to Other Work}
\citeauthor{msgl} (\citeyear{msgl}) introduced SGL for multinomial regression, via the \textit{MSGL} R software \citep{r} package but MSGL is limited in its ability to accommodate over-dispersion. On the other  hand, the R package \textit{MGLM} \citep{kim2018mglm} can perform regularization of overdispersed multinomial models, but not SGL. The non-regularized but overdispersed models available in MGLM are fit via iteratively reweighted Poisson regressions based on the MM algorithm, which require the summations over $p$ and $D$ in the surrogate of the associated log-likelihood to be interchangeable. However, when the SGL is applied to the DM regression model,  per Eq. (\ref{eqn:sgl}), the model parameters in the last term are not separable because the square root and summation operators are not interchangeable \citep{zhang2017regression}.  Therefore, MGLM resorts to proximal gradient descent for regularized multivariate count models despite the greater stability of the MM algorithm for optimization of complex objective functions. That being said, SGL is still not available in MGLM as the penalty does not have an analytic solution in the gradient descent step.  \citeauthor{chen2013variable} (\citeyear{chen2013variable}) proposed fitting the penalized DM regression with a SGL penalty for variable selection but they used block coordinate descent, where the stability of the optimization remains questionable. 

Similar to our proposed methods, others have also leveraged quadratic majorization of penalty functions to simplify optimization. Notably, both \citeauthor{van2011flexible} (\citeyear{van2011flexible}) and \citeauthor{lange2014brief} (\citeyear{lange2014brief}) used a quadratic majorizing function on complex penalties such as the $L_1$ norm or the $L_2$ norm to derive the smooth, convex and separable surrogate functions. \citeauthor{van2011flexible} (\citeyear{van2011flexible}) worked within the context of principal component analysis and \citeauthor{lange2014brief} (\citeyear{lange2014brief}) worked within the context of group lasso regression. More generally, an iterative ridge regression procedure has been recommended for optimizing regression models with $L_q$ penalties for $0 < q \leq 1$, hard-thresholding penalties and the smoothly clipped absolute deviation penalty \citep{fan2001variable, hunter2005variable}. Although our methodology shares similarities with these approaches, our algorithm uniquely consolidates these principles into a unified framework applicable across a diverse range of regularized regression settings with a particular focus on the complex two-term SGL penalty. For generalized linear models and multivariate count models, we provide a simple closed-form solution at each iteration.  

\subsection{Contributions and Outline of the Paper}
This paper has four major contributions to statistical methodology. We: (1) introduce DHR as a novel framework to optimize a penalty function in a regularized regression model; (2) show that our proposed DHR on SGL, lasso, and group lasso penalties can be formulated as an iteratively reweighted ridge regression for distributions in the exponential family; (3) develop a unified framework, the iteratively reweighted Poisson ridge regression algorithm, for optimizing a class of regularized multivariate count regression models; and (4) evaluate the performance of DHR in the context of regularized DM regression. Section 2 introduces the notation and background materials that will be used throughout the paper while Section 3 details the proposed methods. Section 4 evaluates our proposed method in simulation. Section 5 applies DHR in the analysis of benthic macroinvertebrate community data collected from the Athabasca oil sands region. Finally, Section 6 provides conclusions, discussion and future work. It is worth noting that, while our primary focus is on DM regression, we also derive a general solution for optimizing other regularized multivariate count models, including multinomial, negative multinomial, and generalized DM, whose log-likelihoods with regularization are also known to be non-convex, non-smooth, and difficult to optimize.

\section{Notation and Background}
Let $i=1, \ldots, n$ index the observations in a sample of data and $j=1, \ldots, p$ index the covariates. Suppose we have paired data $(y_i, \mathbf{x}_i)$ where $y_i$ is the response and $\mathbf{x}_i$ is the design vector of the $p$ covariates for the $i^{th}$ observation. In what follows, we review several key algorithms that are building blocks for deriving our novel DHR and the subsequent novel algorithm for optimizing the SGL regularized DM regression model.

\subsection{Iteratively Reweighted Least Squares}
A generalized linear model (GLM) for outcome $Y$ with a distribution belonging to the exponential family has a probability distribution function that can be expressed as $f(y; \theta, \phi) = \exp \left[ \frac{y \theta - b(\theta)}{a(\phi)} + c(y, \phi) \right]$, where $\theta$ is the canonical parameter, $\phi$ is the dispersion parameter, and $a(\cdot), b(\cdot), c(\cdot)$ are known functions \citep{FarawayJulianJ2016EtLM}. Let $g(\cdot)$ be a link function that links the linear predictor $\boldsymbol{\eta}=\mathbf{X}\boldsymbol{\beta}$ to the mean of the response $\boldsymbol{\mu}=E(Y|\mathbf{X})$, i.e., $g(\boldsymbol{\mu})=\boldsymbol{\eta}=\mathbf{X}\boldsymbol{\beta}$, where $\boldsymbol{\beta}=(\beta_0, \beta_1,... \beta_p)$ is a vector consisting of the intercept and coefficients associated with each of the $p$ covariates.  The variance function is given by the matrix of second derivatives such that $V(\boldsymbol{\mu}) = b''(\theta)/a(\phi)$.  The MLE of $\boldsymbol{\beta}$ can be obtained via the iteratively reweighted least squares (IRLS) algorithm in which, at iteration $t+1$, we have the closed-form update $\boldsymbol{\beta}^{(t+1)} = \left(\mathbf{X}'\boldsymbol{\Gamma}^{(t)}\mathbf{X}\right)^{-1}\mathbf{X}'\boldsymbol{\Gamma}^{(t)}\mathbf{z}^{(t)}$, where 
\begin{equation}
\begin{split}
\label{eqn:irls_est}
\boldsymbol{\Gamma}^{(t)} &= \left[\left(\left.\left(\frac{\partial \boldsymbol{\eta}}{\partial \boldsymbol{\mu}}\right)^2\right|_{\boldsymbol{\eta}^{(t)}} V(\mu)\right)^{-1}\right] \mbox{and} \\
\mathbf{z}^{(t)} &= \boldsymbol{\eta}^{(t)} + (\boldsymbol{y} - \boldsymbol{\mu}^{(t)})\left.\frac{\partial \boldsymbol{\eta}}{\partial \boldsymbol{\mu}}\right|_{\eta^{(t)}}
\end{split}
\end{equation}
are the $n\times n$ diagonal matrix of weights and $n$-length vector of working responses, respectively. Here, the design matrix $\mathbf{X}$ is of dimension $n\times(p+1)$ with an $n$-vector of 1's in the first column and the observed values of the $p$ covariates in the remaining $p$ columns.

\subsection{Majorization-minimization algorithm}
In order to find the solution that minimizes a target objective function, the MM algorithm begins by selecting a surrogate function that majorizes the objective function and is comparably easier to minimize. The target objective function can be minimized via iterative optimization of the surrogate  \citep{hunter2004tutorial}. For a surrogate function, $g(\theta)$, to majorize the objective function, $f(\theta)$, it must meet the following criteria: 1. $f(\theta^{(t)}) = g(\theta^{(t)}|\theta^{(t)})$ and 2. $f(\theta) \leq g(\theta|\theta^{(t)}), \theta \neq \theta^{(t)}$, where $\theta^{(t)}$ is the value that minimizes $g(\theta|\theta^{(t-1)})$ at the $t^{th}$ iteration. At each iteration of the MM algorithm, we seek to construct the majorizing surrogate $g(\theta|\theta^{(t)})$ to obtain $\theta^{(t+1)}$, which is subsequently used to construct $g(\theta|\theta^{(t+1)})$ for $\theta^{(t+2)}$. This process is repeated until convergence. Given that 
 $f(\theta^{(t+1)}) \leq g(\theta^{(t+1)}|\theta^{(t)}) \leq g(\theta^{(t)}|\theta^{(t)}) = f(\theta^{(t)})$, the MM algorithm is a stable optimization method with non-increasing properties. This property makes the MM algorithm useful for optimizing non-convex objective functions, such as the negative of the log-likelihood of the DM regression model.

To find an appropriate surrogate function, we can use an inequality that leads to a surrogate satisfying the abovementioned two criteria. The dominating hyperplane inequality states that, any convex function $f(\theta)$ that is differentiable can be majorized by a function $g(\theta)$ based on the first order Taylor expansion of $f(\theta)$ about a given point, say $\theta^{(t)}$, such that
\begin{equation}
\label{eqn:dhi}
    g(\theta|\theta^{(t)}) = f(\theta^{(t)}) + f'(\theta^{(t)})(\theta-\theta^{(t)}) \geq f(\theta) \ \forall \ \theta
\end{equation}
and $g(\theta^{(t)}|\theta^{(t)}) = f(\theta^{(t)})$. In our proposed DHR, presented in Section 3.1, we use the dominating hyperplane inequality to find a surrogate function that majorizes $\lambda J(\boldsymbol{\beta})$, the penalty part of our target objective function in Eq. (\ref{eqn:sgl}).

\subsection{Iteratively Reweighted Poisson Regression}
\label{sec:irpr}
We next review IRPR for the optimization of multivariate count models whose log-likelihood functions are generally complex, non-concave, and do not belong to the exponential family, including DM, negative multinomial (NM), and generalized DM regression (GDM) \citep{zhang2017regression}. Suppose we wish to examine the association between covariates $\mathbf{x}_i$ and a $D$-length vector of counts, $\boldsymbol{y}_{i} = (y_{i1}, \ldots, y_{iD})$. The regression parameters for a given multivariate count model such as the multinomial, DM, NM, or GDM regression can be organized in a $(p+1)\times d_e$ matrix, $\mathbf{B}$, where each column, $\mathbf{B}_d$, is a $(p+1)$-length vector and $d_e$ depends on the specific model chosen. We use $\mathbf{B}$ to denote the matrix of regression parameters when, depending on the multivariate count regression model, the matrix could include $\boldsymbol{\beta}$ as well as additional model parameters (see Appendix A). 

\citeauthor{zhang2017regression} (\citeyear{zhang2017regression}) demonstrated that the MLE of $\mathbf{B}$ can be found via an MM algorithm by iteratively finding the value of $\mathbf{B}_d$ that maximizes a surrogate function $g_d(\mathbf{B}_d)$ taking on the form of the log-likelihood function of a weighted Poisson regression. Specifically, in the $(t+1)^{th}$ iteration of the IRPR algorithm, we solve
\begin{equation}
\label{eqn:surrogate2}
    \mathbf{B}^{(t+1)} = \stackrel{\arg \min}{_\mathbf{B}} \sum_{d=1}^{d_e} g_{d}(\mathbf{B}_d|\mathbf{B}_d^{(t)}) + C^{(t)}, 
\end{equation}
where
\begin{equation*}
    g_{d}(\mathbf{B}_d|\mathbf{B}_d^{(t)}) = \sum_{i=1}^{n} \Psi_{id}^{(t)}(-\mu_{id} + y_{id}^{*(t)}\log(\mu_{id})).    
\end{equation*}
Here, $\mu_{id} = \exp(\mathbf{x}_i\mathbf{B}_d)$ is treated as the mean of the $d^{th}$ weighted Poisson regression,  $y_{id}^{*(t)}$ and $\Psi_{id}^{(t)}$ are the working response and weight depending on $\mathbf{B}^{(t)}$, the $\mathbf{B}$ estimates obtained in iteration $t$. This sum of weighted Poisson regressions arises from swapping the summation over the sample size with the summation over the $d_e$ regressions in surrogate of the associated log-likelihood function. For demonstration purposes, we now review the IRPR for DM regression. See Appendix A for details on other multivariate count models. 

Suppose $\mathbf{y} = (y_{1}, y_2, \ldots, y_{D})'$ follows a DM distribution \citep{mosimann1962compound} with positive parameters, $\boldsymbol{\alpha} = (\alpha_{1}, \alpha_{2}, \ldots, \alpha_{D})$. In the case of DM regression, we denote each column of the parameter matrix $\mathbf{B}$ as $\boldsymbol{\beta}_d$ such that $\mathbf{B} = (\boldsymbol{\beta}_1, \ldots, \boldsymbol{\beta}_D)$. The covariates $\mathbf{x}_i$ can be related to the response $\mathbf{y}_i$ through the log-linear function,  $\log(\alpha_{id}) = \mathbf{x}_i\boldsymbol{\beta}_d$. The log-likelihood of the DM regression can be written as:
\begin{align}
\begin{split}
\label{eqn:loglikfunc}
    \ell(\mathbf{B}) &=  \sum_{i=1}^{n} \sum_{d=1}^D c_{id} \sum_{l=0}^{y_{id}-1}\log(\exp(\mathbf{x}_i\boldsymbol{\beta}_d)+l) - \sum_{i=1}^{n} \sum_{l=0}^{y_{i+}-1}\log \left(\sum_{d=1}^D \exp(\mathbf{x}_i\boldsymbol{\beta}_d)+l\right) \\ &+ \sum_{i=1}^{n} \log \left(\frac{y_{i+}!}{y_{i1}! \ldots y_{iD}!}\right),
\end{split}
\end{align}
where $c_{id} = 1$ if $y_{id} > 0$ and 0 otherwise, and $y_{i+} = \sum_{d = 1}^{D} y_{id}$ is the total count for observation $i$. Although the log-likelihood may be concave for certain values of  $\boldsymbol{y}_i$ and $\mathbf{x}_i$, it is not guaranteed to be concave in general \citep{chen2013variable}.

The log-likelihood in Eq. (\ref{eqn:loglikfunc}) can be minorized with the sum of $D$ surrogate functions that can be expressed as the log-likelihood of a weighted Poisson regression maximized via the IRPR algorithm as shown in Eq. (\ref{eqn:surrogate2}). The working response and weight for each observation $i$ at the $(t+1)^{th}$ iteration are given by $y_{id}^{*(t)} = \frac{c_{id}}{\Psi_{id}^{(t)}}\sum_{l=0}^{y_{id}-1}\frac{\exp(\mathbf{x}_i\boldsymbol{\beta}_d^{(t)})}{\exp(\mathbf{x}_i\boldsymbol{\beta}_d^{(t)})+l}$,  and $\Psi_{id}^{(t)} = \sum_{l=0}^{y_{i+}-1}\frac{1}{\sum_{d'=1}^D \exp(\mathbf{x}_i\boldsymbol{\beta}_{d'}^{(t)})+l}$, respectively, for $i=1,\ldots,n$.

\section{Methods}
We now introduce a unifying framework, the \textit{iteratively reweighted Poisson ridge regression}, for optimization of the SGL for a class of multivariate count regression models including multinomial, DM, NM, and GDM regression. Conceptually, \textit{dominating hyperplane regularization} (DHR) 
refers to the surrogate that majorizes the regularizing SGL penalty function. In the majorization step of the MM algorithm, the dominating hyperplane inequality is applied to the SGL penalty to derive this DHR surrogate. In all algorithms and derivations presented, we do not penalize any intercept(s).

\subsection{Dominating Hyperplane Regularization}
We first identify an appropriate majorizing surrogate function for the penalty function in Eq. (\ref{eqn:sgl}) using DHR. To create separability of model parameters in the SGL penalty, we find separate surrogate functions for the $L_1$ and $L_2$ terms via the dominating hyperplane inequality in Eq. (\ref{eqn:dhi}). The two resulting surrogate functions can be combined into a weighted ridge surrogate function in the form of
\begin{equation}
\label{eqn:dhr_surr}    g(\boldsymbol{\beta}|\boldsymbol{\beta}^{(t)}) = \lambda\sum_{k=1}^K
\nu_{k}^{(t)}\beta_{k}^2,
\end{equation}
where $K = \sum_{j=1}^m D_j$ for $m$ groups such that $m\leq p$, and $k \equiv k(j,d)$ is a mapping from the tuple $(j,d)$ for $j=1, \ldots, m$ and $d= 1,\ldots, D_j$ to the set ${1,\ldots, K}$. The ridge weight, $\nu_{k}^{(t)}$, for $k$ corresponding to the subscript of $\beta_{jd}$, is a constant given by 
\begin{equation}
\label{eqn:dhr_weights}
    \nu_{k}^{(t)} \simeq \nu_{jd}^{(t)} = \frac{\alpha}{2\sqrt{ \beta_{jd}^{(t)2}}} + \frac{(1-\alpha)\sqrt{D_j}}{2\sqrt{\sum_{d'=1}^{D_j}\beta_{jd'}^{(t)2}}}.
\end{equation}
For univariate outcome regression models, we use $\nu_k$ for $k = 1, \ldots, K$ such that $K=p$, while for multivariate count regression models, we use $\nu_{jd}$ for $j=1,\ldots, m$ and $d = 1, \ldots, d_e$ such that $m=p$ and $D_j = d_e$. See Appendix C.1 for details. 

Our DHR surrogate for SGL has a form similar to the adaptive SGL as defined in \citeauthor{mendez2021adaptive} \citeyearpar{mendez2021adaptive}, whereas our DHR surrogates for lasso and group lasso have forms similar to the adaptive lasso \citep{zou2006adaptive} and the adaptive group lasso \citep{wang2008note}, respectively. A small quantity $\varepsilon>0$ can be added to the denominator of each term in Eq. (\ref{eqn:dhr_weights}) to avoid division by zero when any elements of  $\boldsymbol{\beta}^{(t)}$ are set to zero. Adding $\varepsilon$ to the denominator will majorize a perturbed version of the objective function which is similar to the original objective function \citep{hunter2005variable}. Alternatively, covariates can be removed from the model once they have values less than some $\varepsilon>0$, which is equivalent to setting the corresponding coefficient to zero but without dividing by zero.

\subsection{Regularized GLM with SGL Penalty}
Suppose we wish to fit a regularized GLM with $1 \leq m \leq p$ groups per Eq. (\ref{eqn:sgl}) via the IRLS algorithm. We propose embedding the weighted ridge surrogate from Eq. (\ref{eqn:dhr_surr}) into the IRLS algorithm, giving rise to an iteratively reweighted ridge regression ($IR^3$) procedure as presented in Algorithm \ref{alg:alg2}. At iteration $t+1$, a solution of $\boldsymbol{\beta}$ is given by
    \begin{equation}
    \label{eqn:update}
        \boldsymbol{\beta}^{(t+1)} = \left(\mathbf{X}'\boldsymbol{\Gamma}^{(t)}\mathbf{X} + \lambda\boldsymbol{\nu}^{(t)}\right)^{-1}\mathbf{X}'\boldsymbol{\Gamma}^{(t)}\mathbf{z}^{(t)}.
    \end{equation}
Here, $\boldsymbol{\Gamma}^{(t)}$ is an $n\times n$ diagonal matrix of weights, $\mathbf{z}^{(t)}$ is an $n$-length vector of working responses per IRLS, and $\boldsymbol{\nu}^{(t)}$ is a conformable diagonal matrix of ridge weights with $(0, \nu_{1}^{(t)}, \nu_{2}^{(t)},\ldots, \nu_{K}^{(t)})$ per Eq. (\ref{eqn:dhr_weights}) along the diagonal. See Appendix C.2 for details. In Example \ref{ex:Pois_ex}, we demonstrate the utility of IR$^3$ for regularized Poisson regression.  

\begin{algorithm}
\caption{Iteratively Reweighted Ridge Regression for optimization of a regularized GLM using SGL penalty.}
\label{alg:alg2}
\begin{algorithmic} 
    \REQUIRE Initial estimates $\boldsymbol{\beta}^{(0)} = (\beta^{(0)}_0, \beta^{(0)}_1, \ldots, \beta^{(0)}_k, \ldots, \beta^{(0)}_K)'$, convergence tolerance, and tuning parameters $\alpha$ and $\lambda$
    \REPEAT
            \STATE Update working weights $\boldsymbol{\Gamma}^{(t)}$ per Eq. (\ref{eqn:irls_est}) 
            \STATE Update working responses $\mathbf{z}^{(t)}$ per Eq. (\ref{eqn:irls_est}) 
            \STATE Update ridge weights $\boldsymbol{\nu}^{(t)} = [0, \nu_1^{(t)}, \ldots, \nu_{K}^{(t)}]$ for $\nu_k^{(t)}$ per Eq. (\ref{eqn:dhr_weights})
            \STATE Update $\boldsymbol{\beta}^{(t+1)} \gets \left(\mathbf{X}'\boldsymbol{\Gamma}^{(t)}\mathbf{X} + \lambda\boldsymbol{\nu}^{(t)}\right)^{-1}\mathbf{X}'\boldsymbol{\Gamma}^{(t)}\mathbf{z}^{(t)}$
        \STATE Set $t \gets t+1$
    \UNTIL convergence of objective function 
    \RETURN $\hat{\boldsymbol{\beta}}$
\end{algorithmic}
\end{algorithm}

\newpage
\begin{example}[IR$^3$ for regularized Poisson Regression]
\label{ex:Pois_ex}
When applying the SGL penalty to a Poisson regression with a log link, we wish to find the value of $\boldsymbol{\beta}$ that minimizes the objective function
\begin{equation*}
\label{eqn:poissgl}
    f(\boldsymbol{\beta}) = -\left[\sum_{i=1}^{n}  y_i\mathbf{x}_i\boldsymbol{\beta} - \exp\left(\mathbf{x}_i\boldsymbol{\beta}\right) - \log\left(y_i!\right)\right] + \alpha\lambda\sum_{j=1}^m \sum_{d=1}^{D_j}|\beta_{jd}| + (1-\alpha)\lambda\sum_{j=1}^m  \sqrt{D_j}   \sqrt{\sum_{d=1}^{D_j}\beta_{jd}^2}.
\end{equation*}
It is known that the mean response at iteration $t$ is given by $\mathbf{\mu}^{(t)} = \exp(\mathbf{X}\boldsymbol{\beta}^{(t)})$. We can use the IR$^3$ with parameter update in Eq. (\ref{eqn:update}) to obtain $\hat{\boldsymbol{\beta}}$ where, per IRLS for Poisson regression, at iteration $t+1$ we have:
\begin{eqnarray*}
\mathbf{z}^{(t)} & = & \mathbf{X}\boldsymbol{\beta}^{(t)} + \left(\boldsymbol{y} -\exp\left(\mathbf{X}\boldsymbol{\beta}^{(t)}\right)\right) \oslash \exp\left(\mathbf{X}\boldsymbol{\beta}^{(t)}\right), \\
\boldsymbol{\Gamma}^{(t)} & = & \text{diag}\left(e^{\mathbf{x}_1^T \boldsymbol{\beta}^{(t)}}, e^{\mathbf{x}_2^T \boldsymbol{\beta}^{(t)}}, \ldots, e^{\mathbf{x}_n^T \boldsymbol{\beta}^{(t)}}\right).
\end{eqnarray*}
The symbol $\oslash$ is used for element-by-element division. The ridge weights $\boldsymbol{\nu}^{(t)}$ are calculated per Eq. (\ref{eqn:dhr_weights}) and are based on the previous iteration's $\beta$'s. We repeat these steps until convergence. 
\end{example}

\subsection{Regularized Multivariate Count Model with SGL}
We now introduce our main result, \textit{iteratively reweighted Poisson ridge regression} (IRPRR), for optimization of regularized multivariate count regression with the SGL penalty. It has been shown in Section \ref{sec:irpr} that a general MM algorithm for finding the MLEs of the regression coefficient parameters $\mathbf{B}$ for multivariate count models including multinomial, DM, NM, and GDM can be formulated as an IRPR procedure. As previously discussed, the SGL penalty naturally extends itself to these multivariate count models given that each row of the matrix $\mathbf{B}$ can be penalized in a group such that there are $p$ groups for $p$ covariates, each comprising $d_e$ regression parameters. Note that the first row of $\mathbf{B}$ comprises the intercepts which are not penalized and therefore we have $p$ groups, not $p+1$. The stable MM algorithm proposed by \citeauthor{zhang2017regression} (\citeyear{zhang2017regression}) cannot be easily applied to the regularized multivariate count model with the SGL penalty due to non-separability of model parameters within the penalty function. Here, we embed the weighted ridge surrogate from Eq. (\ref{eqn:dhr_surr}) into the IRPR algorithm to obtain an IRPRR procedure for optimizing a regularized multivariate count model with the SGL penalty, which is summarized in Algorithm \ref{alg:alg3}. 

At iteration $t+1$, the solution of regression parameters for column $d$ of $\mathbf{B}$ is given by 
    \begin{equation}
    \label{eqn:ridgesol2} \mathbf{B}_d^{\left(t+1\right)} = \left(\mathbf{X}'\mathbf{W}_d^{(t)}\mathbf{X} + \lambda\boldsymbol{\nu}_d^{(t)} \right )^{-1}\mathbf{X}'\mathbf{W}_d^{(t)}\mathbf{z}_d^{(t)},
   \end{equation}
    for $d=1,\ldots,d_e$. Here, $\boldsymbol{\nu}_d^{(t)}$ is a $(p+1)\times (p+1)$ diagonal matrix of ridge weights with $(0, \nu_{1d}^{(t)}, \nu_{2d}^{(t)},\ldots, \nu_{pd}^{(t)})$ along the diagonal per Eq. (\ref{eqn:dhr_weights}). Further, $\mathbf{W}_d^{(t)}$ is an $n\times n$ diagonal matrix with $\Gamma_{id}\Psi_{id}$ on the $i^{th}$ diagonal entry where the weight $\Gamma_{id}^{(t)} = \exp(\mathbf{x}_{i }\mathbf{B}_d^{(t)})$ per IRLS for Poisson regression and the weight $\Psi_{id}$ comes from IRPR and depends on the multivariate count model (see Appendix A).  The $n$-length vector $\mathbf{z}_d^{(t)}$ comprises the working response with $z_{id}^{(t)} = \mathbf{x}_i\mathbf{B}_d^{(t)} + \frac{y_{id}^{*(t)}-\exp\left(\mathbf{x}_i\mathbf{B}_d^{(t)}\right)}{\exp\left(\mathbf{x}_i\mathbf{B}_d^{(t)}\right)}$ for each observation $i$ in which $y_{id}^{*(t)}$ is the working response of the $d^{th}$ regression depending on the multivariate count model in the IRPR algorithm at the $(t+1)^{th}$ iteration. See Appendix C.3 for details. In Example \ref{ex:DM-DHR}, we demonstrate the utility of IRPRR for the SGL regularized DM regression. Appendix B outlines the IRPRR algorithm applied to other multivariate count regression models: regularized multinomial, NM, and GDM regression. 

\begin{algorithm}
\caption{Iteratively Reweighted Poisson Ridge Regression for optimization of regularized multivariate count regression using SGL penalty.}
\label{alg:alg3}
\begin{algorithmic} 
    \REQUIRE Initial estimates $\mathbf{B}^{(0)} = (\mathbf{B}^{(0)}_{1},  \ldots, \mathbf{B}^{(0)}_{d_e})'$, convergence tolerance, and tuning parameters $\alpha$ and $\lambda$
    \REPEAT
    \FOR{$d=1,\ldots,d_e$} 
            \STATE Update working weights $\mathbf{W}_d^{(t)}$ per Appendix B
            \STATE Update working responses $\mathbf{z}_d^{(t)}$ per Appendix B
            \STATE Update ridge weights $\boldsymbol{\nu}_{d}^{(t)} = (0, \nu_{1d}^{(t)}, \ldots, \nu_{pd}^{(t)})$ for $\nu_{jd}^{(t)}$ per Eq. (\ref{eqn:dhr_weights})       
            \STATE Update $\mathbf{B}_d^{(t+1)} \gets \left(\mathbf{X}'\mathbf{W}_d^{(t)}\mathbf{X} + \lambda\boldsymbol{\nu}_d^{(t)}\right)^{-1}\mathbf{X}'\mathbf{W}_d^{(t)}\mathbf{z}_d^{(t)}$
    \ENDFOR
    \STATE Set $t \gets t+1$
    \UNTIL convergence of objective function
    \RETURN $\hat{\mathbf{B}}$
\end{algorithmic}
\end{algorithm}

\begin{example}[IRPRR for regularized Dirichlet-multinomial regression]
\label{ex:DM-DHR}
When regularizing the DM regression using SGL, we wish to find the set of values $\mathbf{B} = (\boldsymbol{\beta}_1, \ldots, \boldsymbol{\beta}_D)$ that minimize the objective function in Eq. (\ref{eqn:sgl}) where $\ell(\mathbf{B})$ is the log-likelihood of the DM regression in Eq. (\ref{eqn:loglikfunc}). At the $\left(t+1\right)^{th}$ iteration of the IRPRR, the solution of $\mathbf{B}^{(t+1)}$ in Eq. (\ref{eqn:ridgesol2}) has the specified $w_{id}^{(t)}$ and $z_{id}^{(t)}$ that make up the weight matrix $\mathbf{W}_d^{(t)}$ and working response vector $\mathbf{z}_d^{(t)}$ for $i=1 \ldots, n$ and $d=1, \ldots, D$, as
\begin{eqnarray*}
    w_{id}^{(t)} & = & \sum_{l=0}^{y_{i+}-1}\frac{\exp\left(\mathbf{x}_{i}\boldsymbol{\beta}_d^{(t)}\right)}{\sum_{d'=1}^D \exp\left(\mathbf{x}_{i}\boldsymbol{\beta}_{d'}^{(t)}\right) + l}, \\
    z_{id}^{(t)} & = &\mathbf{x}_{i}\boldsymbol{\beta}_d^{(t)} + \frac{c_{id}\left(\sum_{l=0}^{y_{id}-1}\frac{\exp\left(\mathbf{x}_{i}\boldsymbol{\beta}_d^{(t)}\right)}{\sum_{d'=1}^D \exp\left(\mathbf{x}_{i}\boldsymbol{\beta}_{d'}^{(t)}\right) + l}\right) -w_{id}^{(t)}}{w_{id}^{(t)}}. 
\end{eqnarray*}

At each iteration in Algorithm 2, we plug the weight matrix, $\mathbf{W}_d^{(t)}$, the working response vector, $\mathbf{z}_d^{(t)}$, and the diagonal matrix of ridge weights $\nu_d^{(t)} = (0, \nu_{1d}^{(t)}, \ldots, \nu_{pd}^{(t)})$ into the step for obtaining $\mathbf{B}_d^{(t+1)}$ until convergence. Algorithm \ref{alg:alg3} in the context of regularized DM regression will herein be referred to as the DM-DHR algorithm.
\end{example}

\subsection{Tuning Parameter Selection}
We select the tuning parameters $\lambda$ and $\alpha$ that minimize the extended Bayesian information criterion (EBIC) \citep{chen2008extended}, defined here as $-2\ell(\boldsymbol{\beta}) + \kappa \log(n) + \kappa \log(K)$,
where $\kappa$ is the number of non-zero coefficients (i.e., $\beta$'s), $n$ is the sample size, and $K$ is the total number of regularized parameters in the model ($p \times D$ in the case of DM regression). Here, the first row of the matrix of regression parameters comprises the intercept terms $\mathbf{B}_{01}, \ldots, \mathbf{B}_{0d_e}$. Since there are an infinite number of possible combinations of $\lambda$ and $\alpha$ values, it is not feasible to evaluate all combinations to identify the optimal one. Therefore, we employ a random search to identify the approximately optimal $\lambda$ and $\alpha$ combination with $\lambda \in [\lambda_{min}, \lambda_{max}]$ and $\alpha \in [0.1, 0.9]$.  In this study, $\lambda_{max}$ is defined as the smallest $\lambda$ that results in the null model without covariates and $\lambda_{min}$ is set as $0.001\lambda_{max}$. Adapting the work of \citeauthor{chen2013variable} (\citeyear{chen2013variable}), we approximate $\lambda_{max}$ by runnning the DM-DHR algorithm along a grid of $\lambda$'s until we reach a $\lambda$ that results in the null model.

\section{Simulation Study}
We evaluated the performance of our novel DM-DHR for SGL through a simulation study adapted from that of \citeauthor{chen2013variable} \citeyearpar{chen2013variable}. In addition, we compared the performance of our DHR algorithm to that of the proximal gradient descent algorithm used in the \textit{MGLM} R package \citep{kim2018mglm} for regularized DM regression with the SGL, lasso, and group lasso penalties. 

\subsection{Simulation Design}
\label{simdesign}
Our simulation design was adapted from that of Chen and Li (2013). Covariates,  $\mathbf{x}_i$, were generated from a multivariate normal distribution with covariance matrix $\Sigma = \{\rho^{|j-k|}\}_{j,k=1}^p$ for $i=1,\ldots,n$, and we set $\rho = 0.4$.  The proportion of relevant covariates was specified by the scalar $\delta_p$ and the proportion of relevant taxa for each covariate remaining in the model was specified by $\delta_D$. The magnitudes of non-zero coefficients were  evenly spaced over the interval $[0.6f, 0.9f]$, where $f$ determines the strength of association such that as $f$ increases, the size of the effect increases. We set $f=0.2$ for weak associations and $f=0.8$ for strong associations. The positive parameters of the DM distribution, $\boldsymbol{\alpha}_i=(\alpha_{i1}, ..., \alpha_{iD})$ were computed via the log-link function $\alpha_{id} = \exp\left(\beta_{0d} + \sum_{j=1}^p \beta_{jd}x_{ij}\right)$. The proportion of the counts, or compositions, $\phi_1, \phi_2, \ldots, \phi_D$, were generated from the Dirichlet($\alpha_1$, $\alpha_2$, \ldots, $\alpha_D$) distribution. The taxa counts $\mathbf{y}_i$ were drawn from a multinomial($\phi_1, \phi_2, \ldots, \phi_D$; $y_{i+}$) distribution where the  total count for observation $i$, $y_{i+} = \sum_{d=1}^D y_{id}$, was drawn from the Poisson distribution with mean equal to 5,000.

\begin{table}
    \centering
        \caption{Simulation study design. Data was generated using each combination of the specified simulation parameter values and replicated 100 times. }
    \begin{tabular}{ll}
    \hline
         \multicolumn{1}{l}{Simulation Parameter} & \multicolumn{1}{l}{Values}  \\
    \hline
         Number of covariates ($p$) & 25, 50, 100 \\
         Number of taxa ($D$) & 7, 12 \\
         Association strength ($f$) & 0.2, 0.8 \\
         \% relevant covariates ($\delta_p$) & 10, 25, 50 \\
         \% relevant taxa per covariate ($\delta_D$) & 25, 50 \\
         Sample size ($n$) & 100, 300, 500 \\
        \hline
         
    \end{tabular}
    \label{tab:sim_design}
\end{table}

The performance of our proposed DM-DHR algorithm was evaluated under settings of different combinations of number of covariates ($p$), number of taxa ($D$), strength of associations ($f$), sample size ($n$), the proportion of relevant covariates ($\delta_p$),
and the proportion of relevant covariate-taxon associations ($\delta_D$). In total, we simulated datasets under a total of 216 different scenarios. For each senario, we generated 100 datasets; see Table \ref{tab:sim_design}. 

To speed up convergence of the DM-DHR algorithm, starting values were obtained from the Broyden-Fletcher-Goldfarb-Shanno (BFGS) algorithm fit to the non-regularized DM regression for a maximum of 20 iterations. When the BFGS algorithm provided unstable estimates, the DM-DHR algorithm would be re-run with starting values of $\mathbf{B}^{(0)}=\mathbf{0}$. When implementing our DM-DHR algorithm, we removed covariates from the design matrix during the estimation step once the respective ridge weights of the group were sufficiently large (i.e., greater than 1e10) to facilitate computational efficiency. 

To evaluate the accuracy of variable selection performance, we used the recall and precision score metrics. These metrics measure variable selection accuracy based on the true positive (TP), false positive (FP), true negative (TN), and false negative (FN) counts per Table 2. For example, coefficients estimated to be non-zero were classified as TP if the corresponding true coefficient was non-zero, and were classified as FP otherwise. The recall and precision score metrics are then defined as:
$\text{recall} = \frac{TP}{TP+FN}$, and  precision $= \frac{TP}{TP+FP}$. As the denominator in recall tallies the total number of true non-zero coefficients, recall represents the discovery rate of relevant covariates. In contrast, as the denominator in precision tallies the total number of coefficients estimated to be non-zero by the model, precision represents the true discovery rate. In this sense, precision is somewhat similar to $1-FDR$, where $FDR$ is the false discovery rate \citep{benjamini1995controlling}.  Both measures take values between 0 and 1 with a score closer to $1$ indicating better variable selection accuracy.

\begin{table}[htp!]
\centering
    \caption{Classification of variables based on whether the associated coefficients were zero in the true underlying data generating process and whether the coefficients were estimated to be zero in the model.}
\begin{tabular}{ccc}
\hline
\multirow{2}{*}{\textbf{True Coefficient}} & \multicolumn{2}{c}{\textbf{Estimated Coefficient}} \\
\cline{2-3}
 & Non-zero & Zero \\
\hline
\multirow{1}{*}{Non-zero} & TP & FN \\
\multirow{1}{*}{Zero} & FP & TN \\
\hline
\end{tabular}
\label{tab:eval_met}
\end{table}

We also measured direction accuracy to determine whether the sign of the estimated coefficients consistently matched that of the true coefficients. Direction accuracy was measured among the true positives and was equal to the percentage of estimated coefficients that had the same sign as the corresponding true coefficient.

\subsection{Results}
For the purpose of brevity and illustration, we only present simulation results for strong association scenarios (i.e., $f=0.8$) in Table \ref{tab:table mm}. Similar patterns were observed for weak associations (i.e., $f=0.2$); however, as expected, lower recall was observed in general compared to the results for $f=0.8$. See Appendix D, Tables D1-D4. 

For the purpose of variable selection, groups corresponded to coefficients associated with one covariate across all taxa. As such, group selection identified covariates associated with at least one taxon, while within-group selection corresponded to identifying specific taxa associated with a given covariate identified in the group selection. Overall, the DM-DHR algorithm demonstrated reasonably high accuracy in identifying true non-zero coefficients for relevant covariate-taxon associations across a diverse range of scenarios as evidenced by its overall mean group recall of 0.882, within-group recall of 0.907, group precision of 0.773 and within-group precision of 0.808 across all scenarios with $f=0.8$. Mean recall and precision remained consistently high across varying levels of the number of taxa ($D$) and varying levels of the proportion of relevant taxa associations ($\delta_D$). Conversely, the ability of the DM-DHR to retain non-zero coefficients was influenced by sample size ($n$), the number of covariates ($p$) and the proportion of relevant covariates ($\delta_p$). For the small sample size ($n=100$), recall decreased as $p$ and $\delta_p$ increased, while for larger sample sizes ($n=300 \text{ or }500$), recall remained high with increasing $p$ and $\delta_p$. For instance, for $n=100$, $p=100$, and $\delta_p=0.5$, the DM-DHR retained only 3\% of the relevant covariates on average. However, when $n=500$, $p=100$, and $\delta_p=0.5$, DM-DHR was able to retain 100\% of relevant covariates on average. 

Precision of the DM-DHR algorithm exhibited less sensitivity to changes in $n$, $p$, and $\delta_p$ compared to recall; although, a slight decline in precision was observed for larger sample sizes ($n=300$ or $n=500$) with increasing $p$ and $\delta_p$ (see Figure 1). This decline in precision was likely due to the precision-recall trade-off given that recall remained high for the larger sample sizes and decreased for the smaller sample size ($n=100$) as $p$ and $\delta_p$ increased. Finally, it is important to highlight the sensitivity of direction accuracy to sample size with increasing $p$ and $\delta_p$. Namely, for $n=100$, a large number of predictors ($p=100$) with a moderate or large proportion of relevant predictors ($\delta_p=0.25, 0.50$), the direction accuracy drops below 50\%. This could be due to the algorithm converging prematurely when the remaining coefficients in the model have negligible influence on the likelihood function, or it could simply be a consequence of the scenarios having $p=n$. When sample size increased to 300 and 500, the direction accuracy remained satisfactory.

\begin{table}[ht!]

\caption{ Mean (SD) Group and within-group selection performance, and direction accuracy of Dirichlet-Multinomial Dominating Hyperplane Regularization for varying sample size ($n$), number of covariates ($p$), and proportion of relevant covariates ($\delta_p$). Values averaged across varying number of taxa ($D=7, 12$) and proportion of relevant taxa ($\delta_D=0.25, 0.5$) and averages across 100 data replicates.}
\label{tab:table mm}
\centering
\fontsize{10}{12}\selectfont
\begin{tabular}[t]{rrrlllll}
\toprule
\multicolumn{3}{c}{ } & \multicolumn{2}{c}{Group Selection} & \multicolumn{2}{c}{Within-group Selection} & \multicolumn{1}{c}{ } \\
\cmidrule(l{3pt}r{3pt}){4-5} \cmidrule(l{3pt}r{3pt}){6-7}
$n$ & $p$ & $\delta_p$ & Precision & Recall & Precision & Recall & Direction acc.\\
\midrule
 &  & 0.10 & 0.84 (0.20) & 1.00 (0.03) & 0.91 (0.10) & 0.89 (0.09) & 0.96 (0.04)\\

 &  & 0.25 & 0.78 (0.15) & 0.98 (0.05) & 0.86 (0.08) & 0.88 (0.06) & 0.97 (0.02)\\

 & \multirow{-5}{*}{\raggedleft\arraybackslash 25} & 0.50 & 0.76 (0.11) & 0.98 (0.11) & 0.80 (0.06) & 0.91 (0.07) & 0.97 (0.05)\\

 &  & 0.10 & 0.82 (0.15) & 0.99 (0.02) & 0.89 (0.08) & 0.73 (0.10) & 0.95 (0.02)\\

 &  & 0.25 & 0.68 (0.13) & 0.99 (0.03) & 0.86 (0.05) & 0.89 (0.06) & 0.97 (0.03)\\

 & \multirow{-5}{*}{\raggedleft\arraybackslash 50} & 0.50 & 0.90 (0.08) & 0.43 (0.27) & 0.82 (0.12) & 0.72 (0.17) & 0.72 (0.17)\\

 &  & 0.10 & 0.80 (0.15) & 0.89 (0.16) & 0.84 (0.06) & 0.81 (0.10) & 0.94 (0.06)\\

 &  & 0.25 & 0.96 (0.10) & 0.07 (0.12) & 0.96 (0.08) & 0.56 (0.20) & 0.47 (0.11)\\

\multirow{-16}{*}[1\dimexpr\aboverulesep+\belowrulesep+\cmidrulewidth]{\raggedleft\arraybackslash 100} & \multirow{-5}{*}{\raggedleft\arraybackslash 100} & 0.50 & 0.99 (0.05) & 0.03 (0.02) & 0.92 (0.14) & 0.46 (0.19) & 0.46 (0.07)\\
\addlinespace
 &  & 0.10 & 0.90 (0.17) & 1.00 (0.00) & 0.80 (0.13) & 0.91 (0.09) & 1.00 (0.00)\\

 &  & 0.25 & 0.80 (0.14) & 1.00 (0.01) & 0.77 (0.07) & 0.99 (0.02) & 1.00 (0.00)\\

 & \multirow{-5}{*}{\raggedleft\arraybackslash 25} & 0.50 & 0.75 (0.11) & 1.00 (0.00) & 0.78 (0.06) & 0.99 (0.01) & 1.00 (0.00)\\

 &  & 0.10 & 0.79 (0.16) & 0.99 (0.02) & 0.83 (0.09) & 0.99 (0.02) & 1.00 (0.00)\\

 &  & 0.25 & 0.73 (0.14) & 1.00 (0.00) & 0.79 (0.06) & 0.98 (0.03) & 1.00 (0.00)\\

 & \multirow{-5}{*}{\raggedleft\arraybackslash 50} & 0.50 & 0.68 (0.07) & 1.00 (0.00) & 0.75 (0.04) & 0.99 (0.01) & 1.00 (0.00)\\

 &  & 0.10 & 0.79 (0.16) & 1.00 (0.01) & 0.85 (0.06) & 0.98 (0.02) & 1.00 (0.00)\\

 &  & 0.25 & 0.63 (0.12) & 0.99 (0.01) & 0.80 (0.06) & 0.97 (0.03) & 1.00 (0.00)\\

\multirow{-16}{*}[1\dimexpr\aboverulesep+\belowrulesep+\cmidrulewidth]{\raggedleft\arraybackslash 300} & \multirow{-5}{*}{\raggedleft\arraybackslash 100} & 0.50 & 0.64 (0.10) & 0.88 (0.12) & 0.71 (0.07) & 0.93 (0.10) & 0.95 (0.06)\\
\addlinespace
 &  & 0.10 & 0.82 (0.21) & 0.99 (0.06) & 0.80 (0.13) & 1.00 (0.03) & 1.00 (0.00)\\

 &  & 0.25 & 0.83 (0.15) & 0.94 (0.22) & 0.79 (0.10) & 0.98 (0.08) & 0.99 (0.04)\\

 & \multirow{-5}{*}{\raggedleft\arraybackslash 25} & 0.50 & 0.71 (0.15) & 0.92 (0.20) & 0.72 (0.08) & 0.98 (0.07) & 0.98 (0.05)\\

 &  & 0.10 & 0.84 (0.16) & 0.96 (0.12) & 0.76 (0.11) & 0.98 (0.07) & 1.00 (0.01)\\

 &  & 0.25 & 0.81 (0.12) & 0.93 (0.17) & 0.79 (0.09) & 0.98 (0.06) & 0.99 (0.03)\\

 & \multirow{-5}{*}{\raggedleft\arraybackslash 50} & 0.50 & 0.66 (0.09) & 0.99 (0.06) & 0.73 (0.06) & 1.00 (0.01) & 1.00 (0.02)\\

 &  & 0.10 & 0.71 (0.12) & 1.00 (0.03) & 0.79 (0.07) & 1.00 (0.01) & 1.00 (0.01)\\

 &  & 0.25 & 0.56 (0.10) & 1.00 (0.00) & 0.79 (0.04) & 1.00 (0.00) & 1.00 (0.00)\\

\multirow{-16}{*}[1\dimexpr\aboverulesep+\belowrulesep+\cmidrulewidth]{\raggedleft\arraybackslash 500} & \multirow{-5}{*}{\raggedleft\arraybackslash 100} & 0.50 & 0.59 (0.04) & 1.00 (0.00) & 0.69 (0.04) & 1.00 (0.00) & 1.00 (0.00)\\
\bottomrule
\end{tabular}
\end{table}

\normalsize

\renewcommand\thesubfigure{\roman{subfigure}} 

\begin{figure}[!ht]
    \centering
    \begin{subfigure}{\linewidth}
        \centering
        \includegraphics[width=.6\linewidth]{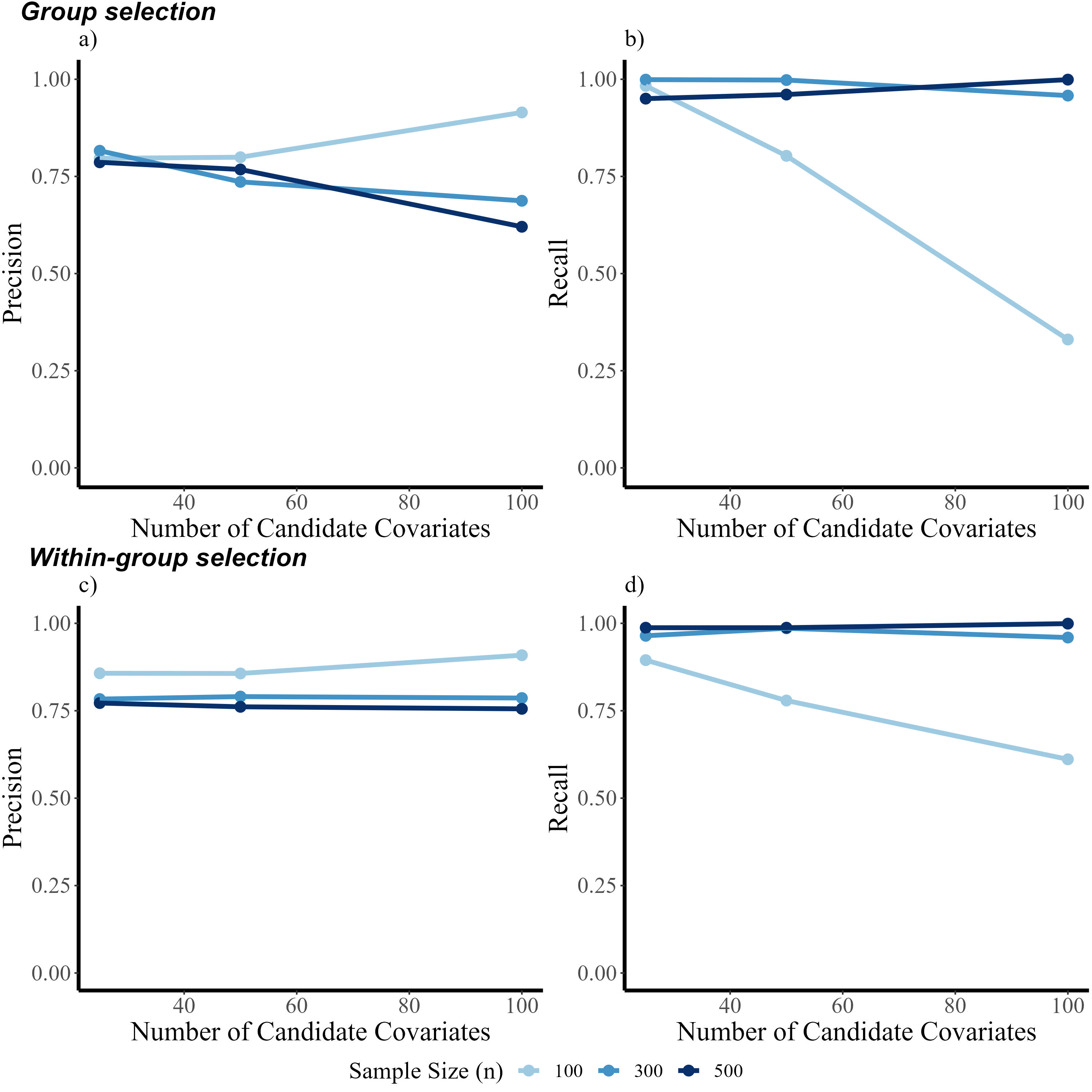}
        \caption{Number of candidate covariates.}
        \label{fig:p_plot}
    \end{subfigure}
    
    \begin{subfigure}{\linewidth}
        \centering
        \includegraphics[width=.6\linewidth]{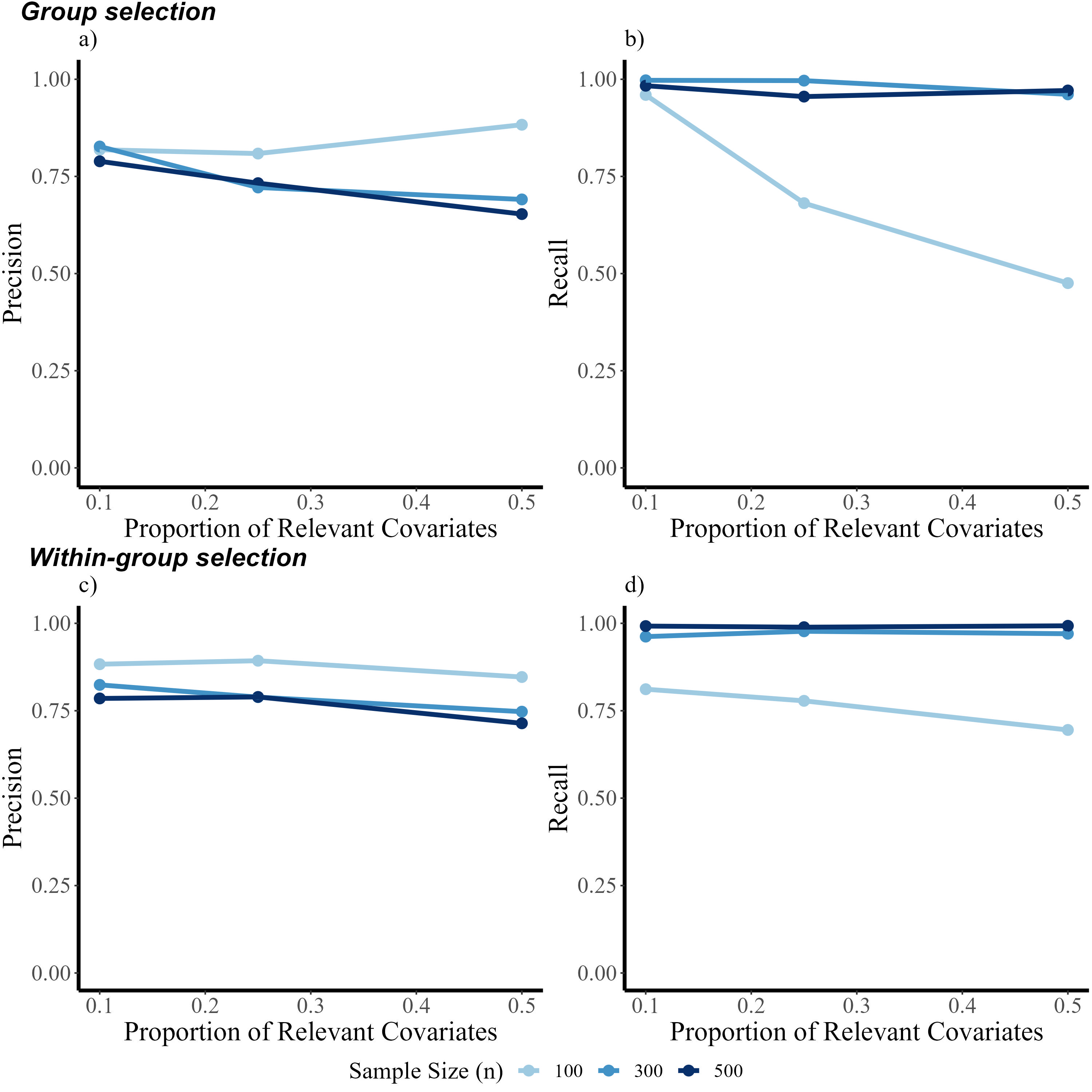}
        \caption{Proportion of relevant covariates.}
        \label{fig:prel_plot}
    \end{subfigure}
    \caption{Precision (left columns) and recall (right columns) by i. number of candidate covariates, ii. proportion of relevant covariates, and by sample size (line colour). Top rows show group selection; bottom rows show within-group selection.}
\end{figure}

Finally, we compared performance of regularized DM regression with the lasso, group, and SGL penalties when implemented via our proposed DM-DHR algorithm versus when implemented via proximal gradient descent within the \textit{MGLM} package in R \citep{kim2018mglm}. Results are presented in Table E1 of the Appendix. DHR and MGLM performed similarly when using the same penalty function. However, DHR proved to be beneficial through its ability to fit the SGL penalty, which demonstrated robust selection performance compared to the lasso and group penalties. For further discussion of these results, we refer the reader to Appendix E.

\section{Application}
\subsection{Data Description}
The Athabasca oil sands region in Canada is home to the world's largest bitumen deposit, and has seen increased industrial activity along the Athabasca River. Environmentalists and stakeholders have raised concern about potential changes in the regional environmental attributes, such as altered water chemistry. For example, saline water discharge from groundwater has likely altered the Athabasca river's chemistry with large increases of chloride concentration \citep{jasechko2012quantifying}. Moreover, concentrations of metals that can be found in bitumen, such as vanadium, nickel, and molybdenum, may become elevated in areas with development activities at levels toxic to local wildlife \citep{bicalho2017determination, kelly2010oil}. Unfortunately, isolating the industrial contribution of these metals from concentrations naturally occurring in the region is not straightforward. Regardless of the specific cause (e.g., natural or industrial), identifying which of these metals may be driving changes in biological communities can help establish a feedback loop to better focus monitoring and research activities in the region for adaptive monitoring programs \citep{arciszewski2017using}.

\begin{figure}[htb!]
    \centering
    \includegraphics[scale=0.65]{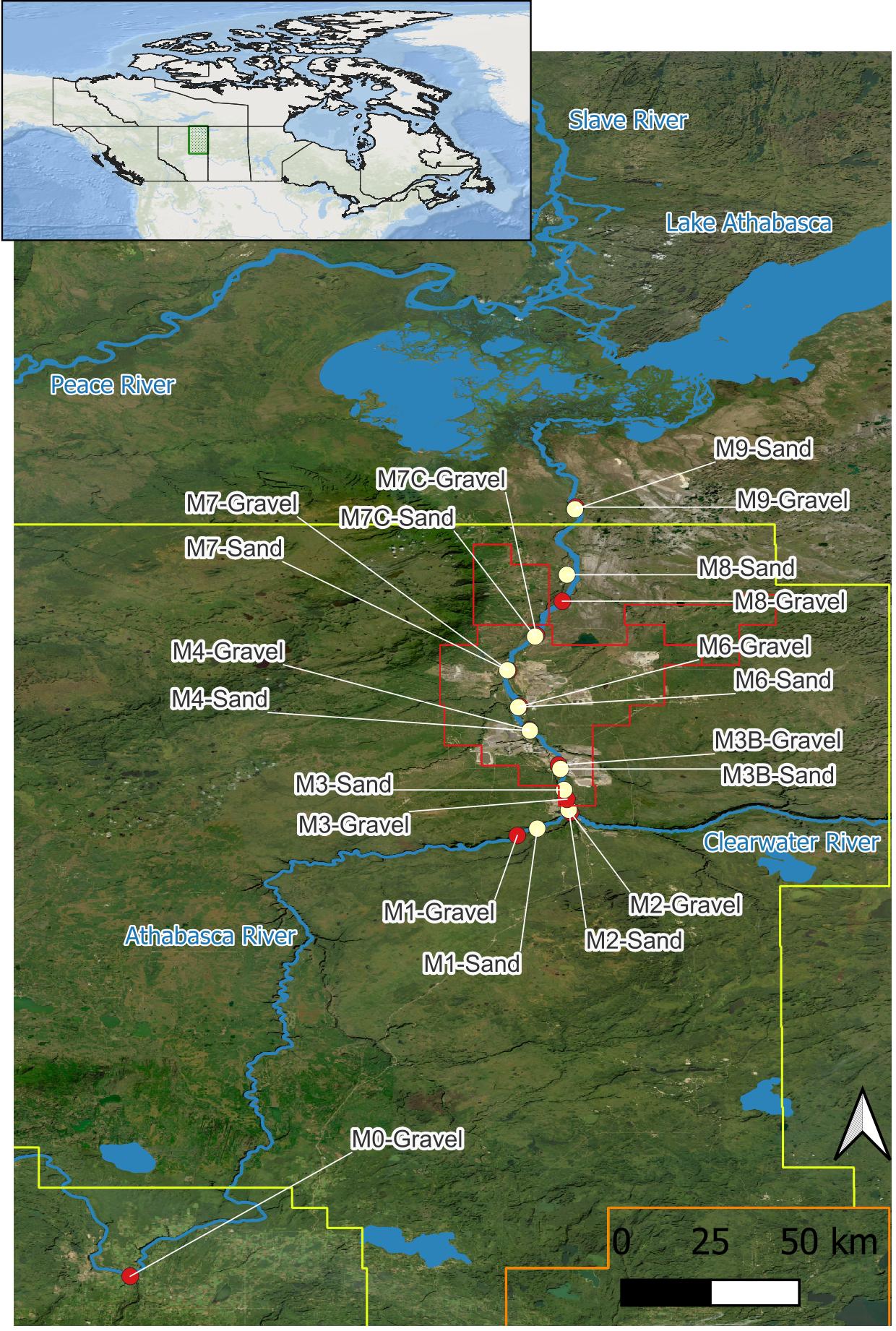}
    \caption{Map of sampling sites for benthic macroinvertebrate along the Athabasca river. Yellow border shows the Athabasca oil sands region, red border shows the minable region, and orange line shows the top of the Cold Lake oil sands region. Basemaps provided by ESRI.}
    \label{fig:sample_sites}
\end{figure}

We obtained data from a study investigating the association between water quality and benthic macroinvertebrate communities inhabiting the Athabasca oil sands region \citep{culp2018assessing}. The data is publicly available at: https://data-donnees.az.ec.gc.ca/data/ substances/monitor/benthic-invertebrates-oil-sands-region/mainstem-benthic-invertebrates-oil-sands-region/. The primary objective of our analysis was to identify potential contaminants of concern using the benthic macroinvertebrate compositions of the Athabasca river as an indicator for the health of the aquatic ecosystem. We focused on identifying metals, nutrients, and ions present in the water that were associated with benthic macroinvertebrate compositions. We included the following ions: calcium, chloride, sodium, phosphorous; metals: magnesium, aluminum, vanadium, nickel, molybdenum, arsenic, cadmium, antimony, cobalt; and nutrients: nitrogen, particulate organic carbon, as covariates in our analysis. We also included an indicator variable for whether the soil substrate was gravel or sand, resulting in a total of $p=16$  covariates. Each observation in the dataset represented a sample from one of 13 sites in a given year from 2012 to 2017, taken from either the gravel or sand, resulting in a total of $n=96$ observations. Figure \ref{fig:sample_sites} presents the sampling sites along the Athabasca river. Benthic macroinvertebrate counts for each observation were aggregated into seven taxa at the rank of order. 

\subsection{Data Analysis}
We fit our DM-DHR method to the benthic community data to identify relevant covariate-taxon associations. We applied three different penalties: the group lasso penalty ($\alpha=0$), the lasso penalty ($\alpha=1$), and the SGL penalty ($0 < \alpha < 1$). For the SGL penalty, $\alpha$ served as an additional tuning parameter taking values in the range of [0.1, 0.9]. To determine the optimal tuning parameter(s) for each model, we conducted a grid search and for each model, selected the tuning parameter values that minimized the EBIC for model fitting. During the grid search, combinations of $\lambda$ and $\alpha$ were evaluated with one hundred $\lambda$ values across a logarithmic scale ranging from $\lambda_{max}$ to $0.001\times\lambda_{max}$ and $\alpha$ set to either 0.1, 0.3, 0.5, 0.7, or 0.9. 

\subsection{Results}

Table \ref{tab:benthic_sg} displays the coefficient estimates for selected covariates obtained through SGL, lasso, and group lasso regularization, respectively. The regression coefficients in red were those additionally selected when relaxing the chosen $\lambda$ to the $\lambda$ that was one grid-point smaller. Regardless of the penalty used, we found positive associations between gravel substrate and macroinvertebrates of order Plecoptera, Tubificida, Trichoptera, Veneroida, and Odonata. However, the SGL model had the lowest EBIC ($6855.909$) compared to lasso (6873.247) and group lasso (6867.507).  In general, the SGL fell between the group lasso and lasso in terms of number of covariates retained in the model (group selection) and specific covariate-taxon associations found (within-group selection).  The group lasso did not identify any associations beyond substrate until the value of $\lambda$ was relaxed, at which point arsenic was retained. Note that the values of estimated non-zero coefficients only changed slightly when relaxing $\lambda$ and, therefore, we did not report these in Table \ref{tab:benthic_sg}.

\begin{table}[ht]
\caption{Coefficient estimates of selected coefficients from DM-DHR applied to benthic macroinvertebrate data at the rank of order with sparse group lasso penalty, lasso penalty, and group lasso penalty. Coefficients in black were selected at the optimal $\lambda$ and coefficients in red were those additionally selected when relaxing $\lambda$ to the previous $\lambda$ on the grid. The other non-zero coefficients when using relaxed lambda are almost identical to those in black and therefore we don't report them separately here. }
\label{tab:benthic_sg}
\fontsize{8}{8}\selectfont
\begin{tabular}[t]{lllllllll}
\toprule
  Penalty & Variable & Diptera & Ephemeroptera & Plecoptera & Tubificida & Trichoptera & Veneroida & Odonata\\
\midrule
SGL & &  &  &  & & & & \\
& Intercept & \textcolor{black}{$\ 1.44$} & \textcolor{black}{$\ 1.36$} & \textcolor{black}{-0.82} & \textcolor{black}{-1.59} & \textcolor{black}{-1.04} & \textcolor{black}{-2.11} & \textcolor{black}{-1.47}\\
& Gravel & \textcolor{black}{$<$ 0.01} & \textcolor{black}{-} & \textcolor{black}{$\ 0.45$} & \textcolor{black}{$\ 0.92$} & \textcolor{black}{$\ 0.88$} & \textcolor{black}{$\ 0.75$} & \textcolor{black}{$\ 0.60$}\\
& Nitrogen & \textcolor{black}{-} & \textcolor{black}{-} & \textcolor{red}{$<0.01$} & \textcolor{black}{-} & \textcolor{black}{-} & \textcolor{black}{-} & \textcolor{black}{-}\\
& Vanadium & \textcolor{black}{-} & \textcolor{black}{-} & \textcolor{black}{-} & \textcolor{red}{$<0.01$} & \textcolor{black}{-} & \textcolor{black}{-} & \textcolor{black}{-}\\
& Arsenic & \textcolor{black}{-} & \textcolor{black}{-} & \textcolor{black}{$\ 0.76$} & \textcolor{black}{$\ 0.90$} & \textcolor{black}{-} & \textcolor{black}{-} & \textcolor{black}{-}\\
\addlinespace
LASSO & &  &  &  & & & & \\
& Intercept & \textcolor{black}{$\ 1.43$} & \textcolor{black}{$\ 1.35$} & \textcolor{black}{-0.78} & \textcolor{black}{-1.56} & \textcolor{black}{-1.02} & \textcolor{black}{-2.04} & \textcolor{black}{-1.43}\\
& Gravel & \textcolor{black}{-} & \textcolor{black}{-} & \textcolor{black}{$\ 0.40$} & \textcolor{black}{$\ 0.88$} & \textcolor{black}{$\ 0.84$} & \textcolor{black}{$\ 0.66$} & \textcolor{black}{$\ 0.52$}\\
& Calcium & \textcolor{black}{-} & \textcolor{black}{-} & \textcolor{black}{-} & \textcolor{black}{-} & \textcolor{black}{-} & \textcolor{black}{-} & \textcolor{red}{$<0.01$}\\
& Nitrogen & \textcolor{black}{-} & \textcolor{black}{-} & \textcolor{black}{$< 0.01$} & \textcolor{black}{-} & \textcolor{black}{-} & \textcolor{black}{-} & \textcolor{black}{-}\\
& Magnesium & \textcolor{black}{-} & \textcolor{black}{-} & \textcolor{black}{-} & \textcolor{black}{-} & \textcolor{black}{-} & \textcolor{black}{-} & \textcolor{red}{-0.01}\\
& Aluminum & \textcolor{black}{-} & \textcolor{black}{-} & \textcolor{black}{-} & \textcolor{black}{$<0.01$} & \textcolor{black}{-} & \textcolor{black}{-} & \textcolor{black}{-}\\
& Arsenic & \textcolor{black}{-} & \textcolor{black}{-} & \textcolor{black}{$\ 0.75$} & \textcolor{black}{ $\ 0.90$} & \textcolor{black}{-} & \textcolor{black}{-} & \textcolor{black}{-}\\

\addlinespace
Group & &  &  &  & & & & \\
LASSO & Intercept & \textcolor{black}{$\  1.62$} & \textcolor{black}{$\ 1.49$} & \textcolor{black}{-0.45} & \textcolor{black}{-1.04} & \textcolor{black}{-0.94} & \textcolor{black}{-2.01} & \textcolor{black}{-1.44}\\
& Gravel & \textcolor{black}{-0.26} & \textcolor{black}{-0.21} & \textcolor{black}{$\ 0.45$} & \textcolor{black}{$\ 0.74$} & \textcolor{black}{$\ 0.69$} & \textcolor{black}{$\ 0.61$} & \textcolor{black}{$\ 0.54$}\\
& Arsenic & \textcolor{red}{-0.09} & \textcolor{red}{$\ 0.08$} & \textcolor{red}{$\ 0.11$} & \textcolor{red}{$ \ 0.12$} & \textcolor{red}{-0.01} & \textcolor{red}{-0.01} & \textcolor{red}{$\ 0.04$}\\
\bottomrule
\end{tabular}
\end{table}

The SGL and lasso models were well aligned, likely because the selected $\alpha$ for SGL was close to one ($\alpha$ = 0.9). Both SGL and lasso found a positive association between arsenic and orders   Plecoptera and Tubificida. The lasso further identified nitrogen-Plecoptera and aluminum-Tubificida associations, but the coefficients were very small. It seems that the inherent group-level sparsity could more effectively zero out these coefficients compared to the lasso penalty. 

\subsection{Implications} 
Identifying key stressors affecting benthic macroinvertebrates in the Athabasca oil sands region facilitates environmental effects monitoring programs that assess important changes to the ecosystem or develop targeted interventions to mitigate adverse effects.  Using DM-DHR, we found that gravel provides a more suitable habitat than sand for most benthic macroinvertebrates. Arsenic, known for its elevated levels in the Athabasca river \citep{culp2020ecological} showed positive associations with Tubificida, which are known to be more tolerant to pollutants compared to other taxa \citep{hall2018analysis, muralidharan2010macroinvertebrates}, and with Plecoptera, which interestingly are known to be sensitive to pollution \citep{muralidharan2010macroinvertebrates}. Since the DM regression models compositions, rather than independent abundances per se, these positive associations may reflect the lower competition encountered by macroinvertebrates of these orders. In other words, our results may reflect that having a higher arsenic concentration creates an environment too harsh for the other orders rather than creating a hospitable one for Tubificida and Plecoptera. 

The other metals, ions and nutrients were not found to be associated with any of the taxa (except for one small taxon-specific association with each of nitrogen and aluminum). This model sparsity is not surprising given that concentrations of contaminants from natural bitumen deposits and mining activity have not surpassed toxicity thresholds in the region as of yet \citep{culp2020ecological}. In addition, metals and/or other stressors may be associated with small particles, which are more abundant in sand than in gravel, so that taxa variability is explained by substrate. Finally, it is also important to consider that our analyses were limited to the measured ions, metals, and nutrients while there could be other water quality indicators that may influence benthic macroinvertebrate compositions.

\section{Discussion}

We introduced dominating hyperplane regularization for stable optimization of objective functions with intricate penalties, as encountered with the SGL. This elegant algorithm is easy to implement and is particularly well-suited for non-smooth, non-convex objective functions, such as regularization of DM, NM, and GDM regression models. While our results were predominately focused on these multivariate count outcome models, DHR can be seamlessly integrated into any regularized regression model featuring the SGL penalty. Our weighted ridge surrogate from DHR facilitates stable optimization and variable selection for diverse applications where the MM algorithm is employed, including survival analysis \citep{hunter2002computing, ding2015new}, DNA sequence analysis \citep{sabatti2002genomewide}, and medical imaging \citep{zhou2024majorization}. We have shown that through DHR, the optimization of the SGL penalty corresponds to an iteratively re-weighted ridge regression. Since the lasso and group lasso are special cases of the SGL, they each can be fitted by an iteratively re-weighted ridge regression as well. The proposed MM algorithm uses weighted penalty factors in the surrogate function that get large as coefficients approach zero thereby shrinking the corresponding coefficients to zero or very close to zero. Through simulation, we demonstrated the DHR algorithm's stability and high precision across diverse settings for regularized DM regression.

In general, we can use DHR to find a surrogate for the penalty function and incorporate it into the IRPR for a multivariate count model.  We showed how one could adjust the weights and working responses in each iteration of the IRPR for any of the multivariate count models in \citeauthor{zhang2017regression} (\citeyear{zhang2017regression}) using our DHR penalty, resulting in the IRPRR algorithm. A thorough analysis of the benthic data comparing the regularized multivariate count models including DM, multinomial, NM, and GDM regression should be conducted in the future.

One limitation to the IRPRR algorithm is that once a coefficient is set to zero, it cannot re-enter the model in future iterations and so, the coefficient is removed. This may lead to a relevant covariate to be removed prematurely from the  model in early iterations. Adding $\varepsilon$ to the denominator of the ridge weights will avoid this problem but does not majorize the original objective function and does not set coefficients directly to zero but instead shrinks them very close to zero \citep{hunter2005variable}. All results reported in this paper used the first method where the coefficient and design matrix were altered at each iteration to remove any coefficients that were very close or equal to zero. However, when re-running scenarios of the simulation study with the latter method, we found the results to be similar.

Future work should explore integrating alternative, faster algorithms within the DHR algorithm when estimation approaches an optimum, as demonstrated in \citeauthor{zhang2017regression} (\citeyear{zhang2017regression}). Alternatively, methods that focus on reducing the number of iterations of the MM algorithm, such as the quasi-Newton method proposed by \citeauthor{zhou2011quasi}, (\citeyear{zhou2011quasi}) or squared iterative methods proposed by \citeauthor{varadhan2008simple} (\citeyear{varadhan2008simple}) may help. 

The selection of tuning parameters $\lambda$ and $\alpha$ was achieved through minimizing the EBIC. We opted for EBIC to prioritize computational efficiency and simpler models while mitigating the risk of false positives. However, this approach is very conservative compared to the more often used cross-validation procedure and may lead to the algorithm's failure to detect relevant associations in scenarios with low power. This limitation was observed in our simulation study where larger sample sizes ($n=300$, or $500$) were required to retain true non-zero coefficients for complex models with a) a large number of candidate predictors, and/or b) a large proportion of relevant predictors. In exploratory work, we found that using a grid search with warm starts improved the recall of the algorithm with some sacrifice to precision. One could also consider alternative tuning parameter selection methods, such as the Pareto front multi-objective function \citep{cattelani2022improved} or the modified L-curve \citep{pei2015modified}, to assess whether more sophisticated approaches would improve performance. If one were to use a cross-validation approach then considerations would have to be made regarding how best to measure the prediction error. For multivariate count models, the prediction error calculation would require the total count to be known such that the multinomial counts can be predicted given the estimated proportions. Alternatively, one could consider using cross-entropy to measure the accuracy of predicted proportions. This is an interesting direction for future research.

While our primary focus was on the SGL penalty, it is worth noting that the DHR algorithm can be applied whenever the penalty function is intricate, provided there exists a suitable surrogate function for optimization. For instance, penalty functions employed in the context of polygenic risk scores often use log penalties, leading to inseparability of model parameters \citep{chen2017prediction}. Applying DHR to these penalties would facilitate the separability of model parameters and simplify optimization.



\bigskip
\begin{center}
{\large\bf SUPPLEMENTARY MATERIAL}
\end{center}

Appendix A reviews the weights and working responses of IRPR and Appendix B details the weights and working responses of IRPRR for several multivariate count models. Appendix C provides the derivation of the DHR surrogate function for the SGL penalty along with derivations of the updates in the $IR^3$ and IRPRR algorithms. Appendix D presents additional simulation results and Appendix E compares selection performance of the DM-DHR method with proximal gradient descent via the MGLM R package for the lasso, group, and sparse group penalties.


\bibliography{Bibliography}       

\newpage
\section*{Supplementary Materials}
\renewcommand{\thetable}{A\arabic{table}}
\setcounter{table}{0}
\appendix

\begin{landscape}
    \section{Iteratively Reweighted Poisson Regression}
    \label{app:irpr}
    \begin{table}[htb!]
        \fontsize{12}{12}\selectfont
        \caption{Model parameters, working weights, and responses at each iteration of the iteratively reweighted Poisson regression for multivariate count models from \citep{zhang2017regression}. All working responses need to be divided by the corresponding working weight. Presented for self-containment.}
        \resizebox{1.4\textheight}{!}{
        \begin{tabular}{llccccc}
        \toprule
        \multicolumn{3}{c}{ } & \multicolumn{2}{c}{$\boldsymbol{\beta}$} & \multicolumn{2}{c}{$\boldsymbol{\alpha}$}  \\
        \cmidrule(l{3pt}r{3pt}){4-5} \cmidrule(l{3pt}r{3pt}){6-7}
     \multicolumn{1}{l}{Model} & 
    \multicolumn{1}{c}{$d_e$} & 
    \multicolumn{1}{c}{$\mathbf{B}$} & 
    \multicolumn{1}{c}{Weight} & 
    \multicolumn{1}{c}{Response} & 
    \multicolumn{1}{c}{Weight} & 
    \multicolumn{1}{c}{Response} \\
        \midrule
            MN & $D-1$ & $\boldsymbol{\beta}_1, \ldots, \boldsymbol{\beta}_{D-1}$ & $y_{i_+}\left(\sum_{d'=1}^{D-1} e^{\mathbf{x}_i \boldsymbol{\beta}_{d'}^{(t)}}\right)^{-1}$ & $y_{id}$ & N/A & N/A \\
            DM & $D$ & $\boldsymbol{\beta}_1, \ldots, \boldsymbol{\beta}_{D}$ & $\left(\sum_{l=0}^{y_{i+}-1}\sum_{d'=1}^D e^{\mathbf{x}_i\boldsymbol{\beta}_{d'}^{(t)}} + l\right)^{-1}$ & $c_{id}\left(\sum_{l=0}^{y_{id}-1}\frac{e^{\mathbf{x}_i\boldsymbol{\beta}_d^{(t)}}}{e^{\mathbf{x}_i\boldsymbol{\beta}_d^{(t)}}+l}\right)$ & N/A & N/A \\
           NM & $D+1$ & $\boldsymbol{\beta}, \boldsymbol{\alpha}_1, \ldots, \boldsymbol{\alpha}_D$ & $\ln\left(\sum_{d'=1}^D e^{\mathbf{x}_i\boldsymbol{\alpha}_{d'}^{(t)}} + 1\right)$ & $ \sum_{l=0}^{y_{i_+}-1}\frac{e^{\mathbf{x}_i\boldsymbol{\beta}^{(t)}}}{e^{\mathbf{x}_i\boldsymbol{\beta}^{(t)}}+l}$ & $\frac{e^{\mathbf{x}_i\boldsymbol{\beta}^{(t+1)}}+y_{i_+}}{\sum_{d'=1}^De^{\mathbf{x}_i\boldsymbol{\alpha}_{d'}^{(t)}}+1}$ & $y_{id}$\\
           GDM & $2(D-1)$ & \multirow{2}{*}{$\boldsymbol{\beta}_1, \ldots, \boldsymbol{\beta}_{D-1}$} & $\sum_{l=0}^{\zeta_{id}-1}\left(e^{\mathbf{x}_i\boldsymbol{\alpha}_d^{(t)}} + e^{\mathbf{x}_i\boldsymbol{\beta}_d^{(t)}} + l\right)^{-1}$ & $c_{i,d+1}\left(\sum_{l=0}^{y_{i,d+1}-1}\frac{e^{\mathbf{x}_i\boldsymbol{\beta}_d^{(t)}}}{e^{\mathbf{x}_i\boldsymbol{\beta}_d^{(t)}}+l}\right)$ & $\sum_{l=0}^{\zeta_{id}-1}\left(e^{\mathbf{x}_i\boldsymbol{\alpha}_d^{(t)}} + e^{\mathbf{x}_i\boldsymbol{\beta}_d^{(t)}} + l\right)^{-1}$ & $c_{id}\left(\sum_{l=0}^{y_{id}-1}\frac{e^{\mathbf{x}_i\boldsymbol{\alpha}_d^{(t)}}}{e^{\mathbf{x}_i\boldsymbol{\alpha}_d^{(t)}}+l}\right)$ \\
           & & \multirow{2}{*}{$\boldsymbol{\alpha}_1, \ldots, \boldsymbol{\alpha}_{D-1}$} & & & & \\
           \\
          \midrule  
        \end{tabular}
        }
        
        \footnotesize{Note: MN$=$multinomial, DM$=$Dirichlet-multinomial, NM$=$negative multinomial, GDM$=$generalized Dirichlet-multinomial.}
        \label{tab:my_label}

    \end{table}

\end{landscape}

\newpage

\section{Iteratively Reweighted Poisson Ridge Regression}
\label{app:irprr}
Below we present the working weights and responses of the IRPRR for the four multivariate count models described in \citeauthor{zhang2017regression} (\citeyear{zhang2017regression}).

\textbf{Regularized multinomial regression}. The multinomial regression has parameters $\mathbf{B} = (\boldsymbol{\beta}_1, \boldsymbol{\beta}_2, \ldots, \boldsymbol{\beta}_D)$ where each $\boldsymbol{\beta}_d$ is a $(p+1)$-length vector. The objective function of the regularized multinomial regression using SGL is,
\begin{eqnarray*}
    -\ell(\mathbf{B}) + \lambda J(\mathbf{B}) & = & -\sum_{d=1}^{D} \sum_{i=1}^{n} y_{id}\left( \mathbf{x}_{i} \boldsymbol{\beta}_d - \ln\sum_{d'=1}^{D} \exp\left(\mathbf{x}_i\boldsymbol{\beta}_{d'}\right)\right) + \sum_{i=1}^n \ln\binom{y_{i_+}} {\mathbf{y}_i} \\ & + & \lambda\alpha\sum_{j=1}^p\sum_{d=1}^{D} |\beta_{jd}| + \lambda (1-\alpha)\sum_{j=1}^p \sqrt{D-1}\sqrt{\sum_{d=1}^{D}\beta_{jd}^2} ,
\end{eqnarray*}
where $\boldsymbol{\beta}_D = \mathbf{0}$ is the reference taxon. In the IRPRR algorithm, the parameters $\boldsymbol{\beta}_1, \boldsymbol{\beta}_2, \ldots, \boldsymbol{\beta}_{D-1}$ are updated via  $D-1$ weighted Poisson ridge regressions with the following working weights and responses,
\begin{eqnarray*}
    w_{id}^{(t)} & = & \frac{\exp\left(\mathbf{x}_{i}\boldsymbol{\beta}_{d}^{(t)}\right)y_{i_+}}{\sum_{d'=1}^{D-1} \exp\left(\mathbf{x}_i \boldsymbol{\beta}_{d'}^{(t)}\right)}, \\ 
    z_{id}^{(t)} & = & \exp\left(\mathbf{x}_i\boldsymbol{\beta}_d^{(t)}\right) + \frac{y_{id}-w_{id}^{(t)}}{w_{id}^{(t)}}
\end{eqnarray*}
for $d=1,\ldots, D-1$.

\textbf{Regularized negative multinomial regression}. Negative multinomial regression has parameters $\mathbf{B} = (\boldsymbol{\beta}, \boldsymbol{\alpha}_1, \boldsymbol{\alpha}_2, \ldots, \boldsymbol{\alpha}_D)$ where $\boldsymbol{\beta}$ and each $\boldsymbol{\alpha}_d$ are $(p+1)$-length vectors. The objective function of the regularized negative multinomial regression using SGL is,
\begin{eqnarray*} 
     -\ell(\mathbf{B}) + \lambda J(\mathbf{B}) & = & -\sum_{i=1}^n \sum_{l=0}^{y_{i_+}-1}\ln\left(\exp\left(\mathbf{x}_i\boldsymbol{\beta}\right) + l\right) - \sum_{i=1}^n \left(\exp\left(\mathbf{x}_i\boldsymbol{\beta}\right)+y_{i_+}\right)\ln\left(\sum_{d=1}^D\exp\left(\mathbf{x}_i\boldsymbol{\alpha}_d\right) + 1\right) \\ & + & \sum_{i=1}^n\sum_{d=1}^D y_{id}\mathbf{x}_i\boldsymbol{\alpha}_d - \sum_{i=1}^n\sum_{d=1}^D\ln y_{id}! \\ & + & \lambda\alpha\sum_{j=1}^p|\beta_j| + \lambda\alpha\sum_{j=1}^p\sum_{d=1}^{D} |\alpha_{jd}| + \lambda (1-\alpha)\sum_{j=1}^p \sqrt{D+1}\sqrt{\beta_{j}^2+\sum_{d=1}^{D}\alpha_{jd}^2}.
\end{eqnarray*}
 In the IRPRR algorithm, $\boldsymbol{\beta}$ is updated via a weighted Poisson ridge regression with the following working weights and responses,
\begin{eqnarray*}
    w_{i}^{(t)} & = & \exp\left(\mathbf{x}_i\boldsymbol{\beta}^{(t)}\right)\ln\left(\sum_{d'=1}^D \exp\left(\mathbf{x}_i\boldsymbol{\alpha}_{d'}^{(t)}\right) + 1\right), \ \text{and} \\  z_{i}^{(t)} & = & \exp\left(\mathbf{x}_i\boldsymbol{\beta}^{(t)}\right) + \frac{\left(\sum_{l=0}^{y_{i_+}-1}\frac{\exp\left(\mathbf{x}_i\boldsymbol{\beta}^{(t)}\right)}{\exp\left(\mathbf{x}_i\boldsymbol{\beta}^{(t)}\right)+l}\right) - w_{i}^{(t)}}{w_{i}^{(t)}}.
\end{eqnarray*}
After obtaining $\boldsymbol{\beta}^{(t+1)}$, $\alpha_1, \alpha_2, \ldots, \alpha_D$ are updated via $D$ weighted Poisson ridge regressions with working weights and responses,
\begin{equation*}
    w_{id}^{(t)} = \exp\left(\mathbf{x}_i\boldsymbol{\alpha}_d^{(t)}\right)\frac{\exp\left(\mathbf{x}_i\boldsymbol{\beta}^{(t+1)}\right)+y_{i_+}}{\sum_{d'=1}^D\exp\left(\mathbf{x}_i\boldsymbol{\alpha}_{d'}^{(t)}\right)+1}, \quad 
    z_{id}^{(t)} = \exp\left(\mathbf{x}_i\boldsymbol{\alpha}_d^{(t)}\right) + \frac{y_{id} - w_{id}^{(t)}}{w_{id}^{(t)}}
\end{equation*}
for $d=1, \ldots, D$.

\textbf{Generalized Dirichlet-multinomial regression}. Generalized Dirichlet-multinomial regression has the parameters $\mathbf{B} = (\boldsymbol{\alpha}_1, \boldsymbol{\alpha}_2, \ldots, \boldsymbol{\alpha}_{D-1}, \boldsymbol{\beta}_1, \boldsymbol{\beta}_2, \ldots, \boldsymbol{\beta}_{D-1})$ where each $\boldsymbol{\beta}_d$ and each $\boldsymbol{\alpha}_d$ are a $p$-length vector. The objective function of the regularized GDM regression using SGL is,
\begin{eqnarray*}
-\ell(\mathbf{B}) + \lambda J(\mathbf{B})  & = & \sum_{i=1}^{n} \sum_{d=1}^{D-1} \left(c_{id} \sum_{l=0}^{y_{id}-1} \ln(\exp\left(\mathbf{x}_i\boldsymbol{\alpha}_d\right) + l) + \sum_{l=0}^{\zeta_{i,d+1}-1}\ln(\exp\left(\mathbf{x}_i\boldsymbol{\beta}_d\right)+l) \right.
 \\ & - & \left. \sum_{l=0}^{\zeta_{i,d}-1}\ln\left(\exp\left(\mathbf{x}_i\boldsymbol{\alpha}_d\right) + \exp\left(\mathbf{x}_i\boldsymbol{\beta}_d\right) +l\right)\right)
 + \sum_{i=1}^{n}\ln \binom{y_{i_+}}{\mathbf{y}_i} + \lambda\alpha\sum_{j=1}^p\sum_{d=1}^{D-1}|\beta_{jd}| \\ & + & \lambda\alpha\sum_{j=1}^p\sum_{d=1}^{D-1} |\alpha_{jd}| + \lambda (1-\alpha)\sum_{j=1}^p \sqrt{2(D-1)}\sqrt{\sum_{d=1}^{D-1}\beta_{jd}^2+\sum_{d=1}^{D-1}\alpha_{jd}^2},
\end{eqnarray*}
where $\zeta_{id} = \sum_{k=d}^D y_{ik}$. In the IRPRR algorithm, the parameters $\alpha_1, \alpha_2, \ldots, \alpha_{D-1}$ are updated via $D-1$ weighted Poisson ridge regressions with the following working weights and responses,
\begin{eqnarray*}
    w_{id}^{(t)} & = &\sum_{l=0}^{\zeta_{id}-1}\frac{\exp\left(\mathbf{x}_i\boldsymbol{\alpha}_{d}\right)}{\left(\exp\left(\mathbf{x}_i\boldsymbol{\alpha}_d^{(t)}\right) + \exp\left(\mathbf{x}_i\boldsymbol{\beta}_d^{(t)}\right) + l \right)},  \ \text{and} \\ z_{id}^{(t)} & = & \exp\left(\mathbf{x}_i\boldsymbol{\alpha}_d^{(t)}\right) + \frac{\left(c_{id}\sum_{l=0}^{y_{id}-1}\frac{\exp\left(\mathbf{x}_i\boldsymbol{\alpha}_d^{(t)}\right)}{\exp\left(\mathbf{x}_i\boldsymbol{\alpha}_d^{(t)}\right)+l}\right)-w_{id}^{(t)}}{w_{id}^{(t)}}.
\end{eqnarray*}
for $d=1, \ldots, D-1$.

The parameters $\beta_1, \beta_2, \ldots, \beta_{D-1}$ are updated by solving $D-1$ weighted Poisson ridge regressions with working weights and responses,
\begin{eqnarray*}
    w_{id}^{(t)} & = & \sum_{l=0}^{\zeta_{id}-1}\frac{\exp\left(\mathbf{x}_i\boldsymbol{\beta}_{d}\right)}{\left(\exp\left(\mathbf{x}_i\boldsymbol{\alpha}_d^{(t)}\right) + \exp\left(\mathbf{x}_i\boldsymbol{\beta}_d^{(t)}\right) + l \right)},  \ \text{and} \\ z_{id}^{(t)} & = & \exp\left(\mathbf{x}_i\boldsymbol{\beta}_d^{(t)}\right) + \frac{\left(c_{i,d+1}\sum_{l=0}^{y_{i,d+1}-1}\frac{\exp\left(\mathbf{x}_i\boldsymbol{\beta}_d^{(t)}\right)}{\exp\left(\mathbf{x}_i\boldsymbol{\beta}_d^{(t)}\right)+l}\right)-w_{id}^{(t)}}{w_{id}^{(t)}}.
\end{eqnarray*}
for $d=1, \ldots, D-1$.

\newpage

\section{Derivations}
\label{app:proofs}
\begin{proof}[C.1 Majorizing surrogate for SGL penalty]
To construct the majorizing function proposed in Eq. (6), we can use the dominating hyperplane inequality  for convex, differentiable functions. First, we replace $|\beta_{jd}|$ with its equivalent form $\sqrt{\beta_{jd}^2}$, and then majorize the $L_1$ and $L_2$ penalty terms by using the dominating hyperplane inequality. We then combine these expansions and simplify the resulting sum. 

In the $L_1$ penalty term, we have
\begin{eqnarray}
    \sum_{j=1}^m\sum_{d=1}^{D_j} |\beta_{jd}| & = & \sum_{j=1}^m\sum_{d=1}^{D_j} \sqrt{\beta_{jd}^2} \\ & \leq & \sum_{j=1}^m\sum_{d=1}^{D_j} \left (\sqrt{\beta_{jd}^{(t)2}} + \frac{\beta_{jd}^2 - \beta_{jd}^{(t)2}}{2\sqrt{\beta_{jd}^{(t)2}}} \right) 
    = \sum_{j=1}^m\sum_{d=1}^{D_j} \left(\frac{|\beta_{jd}^{(t)}|}{2} + \frac{\beta_{jd}^2}{2|\beta_{jd}^{(t)}|}\right) \nonumber. 
\end{eqnarray}
Similarly, in the $L_2$ penalty, we have
\begin{equation} \sum_{j=1}^m\sqrt{D_j}\sqrt{\sum_{d=1}^{D_j}\beta_{jd}^2} \leq \sum_{j=1}^m \sqrt{D_j} \left(\frac{\sqrt{\sum_{d=1}^{D_j}\beta_{jd}^{(t)2}}}{2} + \frac{\sum_{d=1}^{D_j}\beta_{jd}^2}{2\sqrt{\sum_{d=1}^{D_j}\beta_{jd}^{(t)2}}}\right).
\end{equation}
Therefore, the surrogate for the penalty function can be split into two parts, one that does not involve $\beta_{jd}$ and another that does:
\begin{eqnarray*}
    \lambda J(\boldsymbol{\beta}) & \leq & \sum_{j=1}^m\left(\lambda\alpha \sum_{d=1}^{D_j} \frac{|\beta_{jd}^{(t)}|}{2} +  \lambda(1-\alpha)\sqrt{D_j} \frac{\sqrt{\sum_{d=1}^{D_j}\beta_{jd}^{(t)2}}}{2} \right) \\ & + & \sum_{j=1}^m \left(\lambda\alpha\sum_{d=1}^{D_j}\frac{\beta_{jd}^2}{2|\beta_{jd}^{(t)}|} + \lambda(1-\alpha)\sqrt{D_j}\frac{\sum_{d=1}^{D_j}\beta_{jd}^2}{2\sqrt{\sum_{d=1}^{D_j}\beta_{jd}^{(t)2}}}  \right) \nonumber \\
    & = & C^{(t)} + \lambda \sum_{j=1}^m \sum_{d=1}^{D_j} \left(\frac{\alpha}{2|\beta_{jd}^{(t)}|} + \frac{(1-\alpha)\sqrt{D_j}}{2\sqrt{\sum_{d=1}^{D_j}\beta_{jd}^{(t)2}}} \right)\beta_{jd}^2 \nonumber, 
\end{eqnarray*}
and therefore, we see that 
\begin{equation}
\label{eqn:dhr_pen}
 \lambda J(\boldsymbol{\beta}) \leq C^{(t)} + \lambda \sum_{j=1}^m \sum_{d=1}^{D_j}\nu_{jd}^{(t)}\beta_{jd}^2,
\end{equation}
where $C^{(t)} = \frac{J(\boldsymbol{\beta}^{(t)})}{2} = \frac{\lambda}{2} \sum_{j=1}^m\left(\alpha\sum_{d=1}^{D_j}|\beta_{jd}^{(t)}| + (1-\alpha)\sqrt{D_j}\sqrt{\sum_{d=1}^{D_j}\beta_{jd}^{(t)2}}\right)$, and $\nu_{jd}^{(t)} = \frac{\alpha}{2\sqrt{ \beta_{jd}^{(t)2}}} + \frac{(1-\alpha)\sqrt{D_j}}{2\sqrt{\sum_{d'=1}^{D_j}}\beta_{jd'}^{(t)2}}$.
Finally, let $k$ index the sequence of pairs in $(j,d)$ for $j=1,2,\ldots,m$; $d=1,2,\ldots,D_j$. We can then vectorize the $\nu_{jd}^{(t)}$'s and re-write (12) as 
\begin{equation}
\label{eqn:final_dhr_surr}
    \lambda J(\boldsymbol{\beta}) \leq \lambda\sum_{k=1}^K\nu_k^{(t)}\beta_k^2,
\end{equation}
where $K=\sum_{j=1}^mD_j$.
\end{proof}

\begin{proof}[C.2 Parameter update in $IR^3$ algorithm]
    To fit a regularized GLM with the SGL penalty, we seek to iteratively minimize the penalized weighted least squares problem. The IRLS algorithm for finding the MLEs of $\boldsymbol{\beta}$ in an unpenalized GLM at iteration $t+1$ is given by,
    \begin{equation}
        \boldsymbol{\beta}^{(t+1)} = \stackrel{\arg\min}{_{\boldsymbol{\beta}}} \sum_{i=1}^{n} \Gamma^{(t)}_i (z^{(t)}_i - \mathbf{X}_i^{(t)} \boldsymbol{\beta})^2,
    \end{equation}
    where $z_i^{(t)}$ and $\Gamma^{(t)}_i$ are the working response and weights for the $i^{th}$ observation at iteration $t+1$, respectively. Now, suppose we regularize a GLM with the SGL penalty per Eq. (13). At the $(t+1)^{th}$ iteration, we aim to find the solution of $\boldsymbol{\beta}$ that minimizes the objective function,
    \begin{equation}
    \label{eqn:objective2}
        \sum_{i=1}^{n} \Gamma^{(t)}_i (z^{(t)}_i - \mathbf{X}_i^{(t)} \boldsymbol{\beta})^2 + \lambda\sum_{k=1}^K\nu_k^{(t)}\beta_k^2.
    \end{equation}

Setting the derivative of (\ref{eqn:objective2}) with respect to $\boldsymbol{\beta}$ to zero and solving gives the solution of $\boldsymbol{\beta}$ for the $(t+1)$th iteration in matrix form as:
    \begin{equation}
        \boldsymbol{\beta}^{(t+1)} = \left(\mathbf{X}'\boldsymbol{\Gamma}^{(t)}\mathbf{X} + \lambda\boldsymbol{\nu}^{(t)}\right)^{-1}\mathbf{X}'\boldsymbol{\Gamma}^{(t)}\mathbf{z}^{(t)},
    \end{equation}
    which is a weighted ridge solution with $(K+1)\times (K+1)$ diagonal weight matrix $\boldsymbol{\nu}^{(t)}$ in place of the identity matrix. The matrix $\boldsymbol{\Gamma}^{(t)}$ is an $n\times n$ diagonal  matrix with the working weights per IRLS along the diagonal and $\mathbf{z}^{(t)}$ is the $n$-length working response vector per IRLS.
\end{proof}

\begin{proof}[C.3 Parameter update in IRPRR algorithm]
Let $\ell(\mathbf{B})$ be the log-likelihood of the multivariate count model. Suppose we wish to minimize the objective function per Eq. (1). To obtain a solution, we majorize the negative log-likelihood of the multivariate count model with the IRPR (Section 2.3) and apply DHR by majorizing the SGL penalty per Eq. (6). Combining the two surrogate functions provides us with the following surrogate that majorizes the objective function,
\begin{equation*}
       \sum_{d=1}^{d_e} \left(-\sum_{i=1}^{n} \Psi_{id}^{(t)}(-\mu_{id} + y_{id}^{*(t)}\log(\mu_{id}))\right) + \lambda\sum_{j=1}^p\sum_{d=1}^{d_e}v_{jd}^{(t)}B_{jd}^2
\end{equation*}
at iteration $t+1$. Exchanging the summation over $d_e$ with the summation over $p$ in the penalty, we aim to find $\mathbf{B}$ that minimizes the surrogate,
\begin{equation}
\label{eqn:dmdhr_surr}
       \sum_{d=1}^{d_e} \left(-\sum_{i=1}^{n} \Psi_{id}^{(t)}(-\mu_{id} + y_{id}^{*(t)}\log(\mu_{id})) + \lambda\sum_{j=1}^pv_{jd}^{(t)} B_{jd}^2\right),
\end{equation}
which is a series of $d_e$ weighted Poisson ridge regressions.

As the sum of regularized weighted Poisson regressions, the surrogate in Eq. (\ref{eqn:dmdhr_surr}) can be minimized via regularized IRLS. Let $\mathbf{W}_d^{(t)}$ denote the $n\times n$ diagonal working weight matrix with $W_{id} = \{\Gamma_{id}\Psi_{id}\}_{i=1}^n$ along the diagonal where $\Gamma_{id}^{(t)} = \exp(\mathbf{x}_{i}\mathbf{B}_d^{(t)})$ per IRLS for Poisson regression and the weight $\Psi_{id}$ comes from IRPR and depends on the multivariate outcome model (see Appendix \ref{app:irpr}). Further, we define $\mathbf{z}_d^{(t)}$ as the $n$-length vector of working responses with $z_{id} = \mathbf{x}_i\mathbf{B}_d^{(t)} + \frac{y_{id}^{*(t)}-\exp\left(\mathbf{x}_i\mathbf{B}_d^{(t)}\right)}{\exp\left(\mathbf{x}_i\mathbf{B}_d^{(t)}\right)}$ for each observation $i$ in which $y_{id}^{*(t)}$ is the working response of the $d^{th}$ regression in the IRPR algorithm at the $(t+1)^{th}$ iteration (see Appendix \ref{app:irpr}). At iteration $t+1$, we seek to minimize 
\begin{equation*}
   \mathbf{B}_d^{(t+1)}=\stackrel{\arg\min}{_{\boldsymbol{\beta}_d}} \left(\mathbf{z}_d^{(t)} - \boldsymbol{\mu}_d^{(t)}\right)'\mathbf{W}_d^{(t)}\left(\mathbf{z}_d^{(t)} - \boldsymbol{\mu}_d^{(t)}\right) + \lambda\sum_{j=1}^p\nu_{jd}^{(t)}B_{jd}^2,
\end{equation*}
which can be solved using the update from Eq. (8),
\begin{equation}
    \mathbf{B}_d^{(t+1)} = \left(\mathbf{X}'\mathbf{W}_d^{(t)}\mathbf{X} + \boldsymbol{\nu}_d^{(t)}\lambda \right )^{-1}\mathbf{X}'\mathbf{W}_d^{(t)}\mathbf{z}_d^{(t)},
\end{equation}
for $d=1,\ldots,d_e$. Here, $\boldsymbol{\nu}_d^{(t)}$ represents the $(p+1) \times (p+1)$ diagonal matrix with $(0, \nu_{1d}, \ldots, \nu_{pd})$ along the diagonal.
\end{proof}

\newpage
\renewcommand{\thetable}{D\arabic{table}}
\setcounter{table}{0}
\section{Simulation Results}
\label{app:results}

\begin{table}[htp!]

\caption{Mean (SD) group and within-group selection performance of the Dirichlet-multinomial Dominating Hyperplane Regularization algorithm for simulation scenarios with varying levels of association strength ($f$), sample size ($n$), and number of covariates ($p$). Values averaged across varying number of taxa ($D = 7, 12$), proportion of relevant taxa ($\delta_D = 0.25, 0.5$), and proportion of relevant covariates ($\delta_p = 0.1, 0.25, 0.5$) and averaged across 100 data replicates.}
\label{tab:table a1}
\centering
\fontsize{10}{12}\selectfont
\begin{tabular}[t]{rrrlllll}
\toprule
\multicolumn{3}{c}{ } & \multicolumn{2}{c}{Group Selection} & \multicolumn{2}{c}{Within-group Selection} & \multicolumn{1}{c}{ } \\
\cmidrule(l{3pt}r{3pt}){4-5} \cmidrule(l{3pt}r{3pt}){6-7}
f & n & p & Precision & Recall & Precision & Recall & Direction accuracy\\
\midrule
 &  & 25 & 0.99 (0.03) & 0.01 (0.05) & 0.94 (0.06) & 0.52 (0.11) & 0.90 (0.06)\\

 &  & 50 & 0.92 (0.15) & 0.01 (0.03) & 0.97 (0.06) & 0.39 (0.08) & 0.88 (0.04)\\

 & \multirow{-4}{*}[1\dimexpr\aboverulesep+\belowrulesep+\cmidrulewidth]{\raggedleft\arraybackslash 100} & 100 & 1.00 (0.00) & 0.00 (0.01) & 1.00 (0.00) & 0.34 (0.04) & 0.85 (0.03)\\

 &  & 25 & 0.94 (0.11) & 0.37 (0.27) & 0.93 (0.11) & 0.49 (0.07) & 1.00 (0.00)\\

 &  & 50 & 0.97 (0.06) & 0.19 (0.21) & 0.95 (0.07) & 0.47 (0.09) & 0.99 (0.01)\\

 & \multirow{-4}{*}[1\dimexpr\aboverulesep+\belowrulesep+\cmidrulewidth]{\raggedleft\arraybackslash 300} & 100 & 0.99 (0.03) & 0.06 (0.07) & 0.99 (0.04) & 0.45 (0.07) & 0.93 (0.02)\\

 &  & 25 & 0.96 (0.07) & 0.50 (0.15) & 0.91 (0.09) & 0.53 (0.11) & 1.00 (0.00)\\

 &  & 50 & 0.98 (0.06) & 0.43 (0.18) & 0.91 (0.08) & 0.51 (0.10) & 0.99 (0.01)\\

\multirow{-13.1}{*}[4\dimexpr\aboverulesep+\belowrulesep+\cmidrulewidth]{\raggedleft\arraybackslash 0.2} & \multirow{-4}{*}[1\dimexpr\aboverulesep+\belowrulesep+\cmidrulewidth]{\raggedleft\arraybackslash 500} & 100 & 0.99 (0.03) & 0.30 (0.16) & 0.94 (0.06) & 0.44 (0.07) & 0.97 (0.02)\\
\addlinespace
 &  & 25 & 0.80 (0.15) & 0.98 (0.06) & 0.86 (0.08) & 0.89 (0.08) & 0.96 (0.04)\\

 &  & 50 & 0.80 (0.12) & 0.80 (0.11) & 0.86 (0.08) & 0.78 (0.11) & 0.88 (0.07)\\

 & \multirow{-4}{*}[1\dimexpr\aboverulesep+\belowrulesep+\cmidrulewidth]{\raggedleft\arraybackslash 100} & 100 & 0.91 (0.10) & 0.33 (0.10) & 0.91 (0.09) & 0.61 (0.16) & 0.62 (0.08)\\

 &  & 25 & 0.82 (0.14) & 1.00 (0.00) & 0.78 (0.09) & 0.96 (0.04) & 1.00 (0.00)\\

 &  & 50 & 0.74 (0.13) & 1.00 (0.01) & 0.79 (0.06) & 0.99 (0.02) & 1.00 (0.00)\\

 & \multirow{-4}{*}[1\dimexpr\aboverulesep+\belowrulesep+\cmidrulewidth]{\raggedleft\arraybackslash 300} & 100 & 0.69 (0.13) & 0.96 (0.05) & 0.79 (0.06) & 0.96 (0.05) & 0.98 (0.02)\\

 &  & 25 & 0.79 (0.17) & 0.95 (0.16) & 0.77 (0.10) & 0.99 (0.06) & 0.99 (0.03)\\

 &  & 50 & 0.77 (0.12) & 0.96 (0.12) & 0.76 (0.09) & 0.99 (0.05) & 0.99 (0.02)\\

\multirow{-13.1}{*}[4\dimexpr\aboverulesep+\belowrulesep+\cmidrulewidth]{\raggedleft\arraybackslash 0.8} & \multirow{-4}{*}[1\dimexpr\aboverulesep+\belowrulesep+\cmidrulewidth]{\raggedleft\arraybackslash 500} & 100 & 0.62 (0.09) & 1.00 (0.01) & 0.76 (0.05) & 1.00 (0.00) & 1.00 (0.00)\\
\bottomrule
\end{tabular}
\end{table}

\begin{table}
\caption{Mean (SD) group and within-group selection performance of the Dirichlet-multinomial Dominating Hyperplane Regularization algorithm for simulation scenarios with varying levels of association strength ($f$), sample size ($n$), and proportion of relevant covariates ($\delta_p$). Values averaged across varying number of taxa ($D = 7, 12$), number of candidate predictors ($p = 25, 50, 100$), and proportion of relevant taxa ($\delta_D = 0.25, 0.5$) and averaged across 100 data replicates.}
\label{tab:table a2}
\centering
\fontsize{10}{12}\selectfont
\begin{tabular}[t]{rrrlllll}
\toprule
\multicolumn{3}{c}{ } & \multicolumn{2}{c}{Group Selection} & \multicolumn{2}{c}{Within-group Selection} & \multicolumn{1}{c}{ } \\
\cmidrule(l{3pt}r{3pt}){4-5} \cmidrule(l{3pt}r{3pt}){6-7}
f & n & $\delta_p$ & Precision & Recall & Precision & Recall & Direction accuracy\\
\midrule
 &  & 0.10 & 0.92 (0.14) & 0.01 (0.05) & 0.95 (0.02) & 0.46 (0.11) & 0.91 (0.06)\\

 &  & 0.25 & 0.99 (0.03) & 0.01 (0.03) & 0.96 (0.07) & 0.37 (0.07) & 0.87 (0.04)\\

 & \multirow{-4}{*}[1\dimexpr\aboverulesep+\belowrulesep+\cmidrulewidth]{\raggedleft\arraybackslash 100} & 0.50 & 1.00 (0.00) & 0.00 (0.01) & 1.00 (0.00) & 0.43 (0.03) & 0.85 (0.02)\\

 &  & 0.10 & 0.93 (0.12) & 0.31 (0.26) & 0.95 (0.10) & 0.49 (0.08) & 1.00 (0.00)\\

 &  & 0.25 & 0.98 (0.05) & 0.16 (0.17) & 0.95 (0.08) & 0.49 (0.09) & 0.99 (0.01)\\

 & \multirow{-4}{*}[1\dimexpr\aboverulesep+\belowrulesep+\cmidrulewidth]{\raggedleft\arraybackslash 300} & 0.50 & 0.99 (0.02) & 0.16 (0.13) & 0.96 (0.05) & 0.43 (0.08) & 0.93 (0.02)\\

 &  & 0.10 & 0.98 (0.07) & 0.33 (0.19) & 0.92 (0.10) & 0.46 (0.09) & 1.00 (0.01)\\

 &  & 0.25 & 0.97 (0.05) & 0.49 (0.15) & 0.92 (0.06) & 0.54 (0.11) & 0.99 (0.01)\\

\multirow{-13.1}{*}[4\dimexpr\aboverulesep+\belowrulesep+\cmidrulewidth]{\raggedleft\arraybackslash 0.2} & \multirow{-4}{*}[1\dimexpr\aboverulesep+\belowrulesep+\cmidrulewidth]{\raggedleft\arraybackslash 500} & 0.50 & 0.98 (0.04) & 0.41 (0.15) & 0.91 (0.07) & 0.47 (0.08) & 0.98 (0.02)\\
\addlinespace
 &  & 0.10 & 0.82 (0.17) & 0.96 (0.07) & 0.88 (0.08) & 0.81 (0.10) & 0.95 (0.04)\\

 &  & 0.25 & 0.81 (0.13) & 0.68 (0.07) & 0.89 (0.07) & 0.78 (0.11) & 0.80 (0.05)\\

 & \multirow{-4}{*}[1\dimexpr\aboverulesep+\belowrulesep+\cmidrulewidth]{\raggedleft\arraybackslash 100} & 0.50 & 0.88 (0.08) & 0.48 (0.13) & 0.85 (0.11) & 0.70 (0.14) & 0.72 (0.09)\\

 &  & 0.10 & 0.83 (0.16) & 1.00 (0.01) & 0.82 (0.09) & 0.96 (0.05) & 1.00 (0.00)\\

 &  & 0.25 & 0.72 (0.14) & 1.00 (0.01) & 0.79 (0.06) & 0.98 (0.02) & 1.00 (0.00)\\

 & \multirow{-4}{*}[1\dimexpr\aboverulesep+\belowrulesep+\cmidrulewidth]{\raggedleft\arraybackslash 300} & 0.50 & 0.69 (0.09) & 0.96 (0.04) & 0.75 (0.06) & 0.97 (0.04) & 0.98 (0.02)\\

 &  & 0.10 & 0.79 (0.17) & 0.98 (0.07) & 0.79 (0.10) & 0.99 (0.04) & 1.00 (0.01)\\

 &  & 0.25 & 0.73 (0.12) & 0.96 (0.13) & 0.79 (0.08) & 0.99 (0.05) & 0.99 (0.02)\\

\multirow{-13.1}{*}[4\dimexpr\aboverulesep+\belowrulesep+\cmidrulewidth]{\raggedleft\arraybackslash 0.8} & \multirow{-4}{*}[1\dimexpr\aboverulesep+\belowrulesep+\cmidrulewidth]{\raggedleft\arraybackslash 500} & 0.50 & 0.65 (0.09) & 0.97 (0.09) & 0.71 (0.06) & 0.99 (0.03) & 0.99 (0.02)\\
\bottomrule
\end{tabular}
\end{table}

\begin{table}
\caption{Mean (SD) group and within-group selection performance of the Dirichlet-multinomial Dominating Hyperplane Regularization algorithm for simulation scenarios with varying levels of association strength ($f$), sample size ($n$), and number of taxa ($D$). Values averaged varying number of candidate predictors ($p = 25, 50, 100$), proportion of relevant covariates ($\delta_p = 0.1, 0.25, 0.5$), and proportion of relevant taxa ($\delta_D = 0.25, 0.5$) and averaged across 100 data replicates.}
\label{tab:table a3}
\centering
\fontsize{10}{12}\selectfont
\begin{tabular}[t]{rrrlllll}
\toprule
\multicolumn{3}{c}{ } & \multicolumn{2}{c}{Group Selection} & \multicolumn{2}{c}{Within-group Selection} & \multicolumn{1}{c}{ } \\
\cmidrule(l{3pt}r{3pt}){4-5} \cmidrule(l{3pt}r{3pt}){6-7}
f & n & D & Precision & Recall & Precision & Recall & Direction accuracy\\
\midrule
 &  & 7 & 0.95 (0.10) & 0.01 (0.03) & 0.97 (0.03) & 0.54 (0.08) & 0.88 (0.04)\\

 & \multirow{-2.5}{*}[0.5\dimexpr\aboverulesep+\belowrulesep+\cmidrulewidth]{\raggedleft\arraybackslash 100} & 12 & 0.99 (0.02) & 0.01 (0.03) & 0.97 (0.04) & 0.30 (0.06) & 0.87 (0.04)\\

 &  & 7 & 0.96 (0.08) & 0.24 (0.20) & 0.96 (0.07) & 0.57 (0.09) & 0.98 (0.01)\\

 & \multirow{-2.5}{*}[0.5\dimexpr\aboverulesep+\belowrulesep+\cmidrulewidth]{\raggedleft\arraybackslash 300} & 12 & 0.97 (0.06) & 0.18 (0.17) & 0.95 (0.08) & 0.37 (0.08) & 0.96 (0.01)\\

 &  & 7 & 0.99 (0.04) & 0.19 (0.16) & 0.95 (0.08) & 0.49 (0.09) & 0.99 (0.00)\\

\multirow{-8.5}{*}[2.5\dimexpr\aboverulesep+\belowrulesep+\cmidrulewidth]{\raggedleft\arraybackslash 0.2} & \multirow{-2.5}{*}[0.5\dimexpr\aboverulesep+\belowrulesep+\cmidrulewidth]{\raggedleft\arraybackslash 500} & 12 & 0.96 (0.07) & 0.63 (0.17) & 0.88 (0.07) & 0.49 (0.10) & 0.98 (0.02)\\
\addlinespace
 &  & 7 & 0.85 (0.12) & 0.67 (0.11) & 0.89 (0.09) & 0.77 (0.13) & 0.81 (0.07)\\

 & \multirow{-2.5}{*}[0.5\dimexpr\aboverulesep+\belowrulesep+\cmidrulewidth]{\raggedleft\arraybackslash 100} & 12 & 0.82 (0.13) & 0.74 (0.07) & 0.86 (0.08) & 0.75 (0.10) & 0.83 (0.06)\\

 &  & 7 & 0.75 (0.14) & 0.97 (0.04) & 0.81 (0.07) & 0.98 (0.03) & 0.99 (0.01)\\

 & \multirow{-2.5}{*}[0.5\dimexpr\aboverulesep+\belowrulesep+\cmidrulewidth]{\raggedleft\arraybackslash 300} & 12 & 0.74 (0.12) & 1.00 (0.00) & 0.76 (0.07) & 0.96 (0.04) & 1.00 (0.00)\\

 &  & 7 & 0.72 (0.13) & 0.98 (0.08) & 0.78 (0.08) & 0.99 (0.03) & 1.00 (0.01)\\

\multirow{-8.5}{*}[2.5\dimexpr\aboverulesep+\belowrulesep+\cmidrulewidth]{\raggedleft\arraybackslash 0.8} & \multirow{-2}{*}[0.5\dimexpr\aboverulesep+\belowrulesep+\cmidrulewidth]{\raggedleft\arraybackslash 500} & 12 & 0.73 (0.13) & 0.96 (0.11) & 0.75 (0.08) & 0.99 (0.04) & 0.99 (0.03)\\
\bottomrule
\end{tabular}
\end{table}

\begin{table}
\caption{Mean (SD) group and within-group selection performance of the Dirichlet-multinomial Dominating Hyperplane Regularization algorithm for simulation scenarios with varying levels of association strength ($f$), sample size ($n$), and proportion of relevant covariate-taxon associations ($\delta_D$). Values averaged across varying number of candidate predictors ($p = 25, 50, 100$), proportion of relevant covariates ($\delta_p = 0.1, 0.25, 0.5$), and number of taxa ($D = 7, 12$) and averaged across 100 data replicates.}
\label{tab:table a4}
\centering
\fontsize{10}{12}\selectfont
\begin{tabular}[t]{rrrlllll}
\toprule
\multicolumn{3}{c}{ } & \multicolumn{2}{c}{Group Selection} & \multicolumn{2}{c}{Within-group Selection} & \multicolumn{1}{c}{ } \\
\cmidrule(l{3pt}r{3pt}){4-5} \cmidrule(l{3pt}r{3pt}){6-7}
f & n & $\delta_d$ & Precision & Recall & Precision & Recall & Direction accuracy\\
\midrule
 &  & 0.25 & 0.95 (0.09) & 0.01 (0.03) & 0.98 (0.04) & 0.51 (0.07) & 0.89 (0.05)\\

 & \multirow{-2.5}{*}[0.5\dimexpr\aboverulesep+\belowrulesep+\cmidrulewidth]{\raggedleft\arraybackslash 100} & 0.50 & 0.99 (0.02) & 0.01 (0.03) & 0.96 (0.03) & 0.31 (0.08) & 0.87 (0.04)\\

 &  & 0.25 & 0.95 (0.09) & 0.23 (0.21) & 0.94 (0.10) & 0.58 (0.08) & 0.99 (0.00)\\

 & \multirow{-2.5}{*}[0.5\dimexpr\aboverulesep+\belowrulesep+\cmidrulewidth]{\raggedleft\arraybackslash 300} & 0.50 & 0.98 (0.05) & 0.19 (0.16) & 0.97 (0.06) & 0.36 (0.09) & 0.95 (0.02)\\

 &  & 0.25 & 0.98 (0.05) & 0.34 (0.16) & 0.89 (0.10) & 0.56 (0.09) & 0.99 (0.01)\\

\multirow{-8.5}{*}[2.5\dimexpr\aboverulesep+\belowrulesep+\cmidrulewidth]{\raggedleft\arraybackslash 0.2} & \multirow{-2.5}{*}[0.5\dimexpr\aboverulesep+\belowrulesep+\cmidrulewidth]{\raggedleft\arraybackslash 500} & 0.50 & 0.97 (0.05) & 0.48 (0.17) & 0.94 (0.05) & 0.42 (0.09) & 0.98 (0.01)\\
\addlinespace
 &  & 0.25 & 0.86 (0.12) & 0.74 (0.11) & 0.85 (0.10) & 0.81 (0.11) & 0.84 (0.07)\\

 & \multirow{-2.5}{*}[0.5\dimexpr\aboverulesep+\belowrulesep+\cmidrulewidth]{\raggedleft\arraybackslash 100} & 0.50 & 0.81 (0.13) & 0.67 (0.07) & 0.90 (0.08) & 0.71 (0.13) & 0.81 (0.06)\\

 &  & 0.25 & 0.78 (0.12) & 1.00 (0.01) & 0.76 (0.08) & 0.96 (0.04) & 1.00 (0.00)\\

 & \multirow{-2.5}{*}[0.5\dimexpr\aboverulesep+\belowrulesep+\cmidrulewidth]{\raggedleft\arraybackslash 300} & 0.50 & 0.71 (0.14) & 0.97 (0.03) & 0.81 (0.06) & 0.98 (0.03) & 0.99 (0.01)\\

 &  & 0.25 & 0.79 (0.13) & 0.96 (0.12) & 0.71 (0.10) & 0.99 (0.04) & 0.99 (0.02)\\

\multirow{-8.5}{*}[2.5\dimexpr\aboverulesep+\belowrulesep+\cmidrulewidth]{\raggedleft\arraybackslash 0.8} & \multirow{-2.5}{*}[0.5\dimexpr\aboverulesep+\belowrulesep+\cmidrulewidth]{\raggedleft\arraybackslash 500} & 0.50 & 0.66 (0.13) & 0.98 (0.07) & 0.81 (0.06) & 0.99 (0.03) & 1.00 (0.02)\\
\bottomrule
\end{tabular}
\end{table}

\clearpage
\renewcommand{\thetable}{E\arabic{table}}
\setcounter{table}{0}
\section{Comparison with Others}
\label{app:comparison}

We compared the performance of regularized DM regression when implemented via our proposed DM-DHR versus when implemented via proximal gradient descent within the \textit{MGLM} package in R \citep{kim2018mglm} for the SGL, lasso, and group lasso penalties. Note that, MGLM only supports lasso and group lasso penalties and does not offer SGL. For the purpose of this comparative analysis, we generated 100 data replicates characterized by sample size ($n=300$), number of taxa ($D=12$), within-group-level sparsity ($\delta_D=0.25$), differing group-level sparsity ($\delta_p=0.1, 0.25, 0.5$), and differing numbers of candidate predictors ($p=25,50,100$). 

The results, presented in Table E1, present the group and within-group selection performance along with direction accuracy for regularized DM regression with the lasso, SGL, and group lasso penalties fitted with either DHR or MGLM. Overall, the SGL penalty offers robust performance across scenarios for both group- and within-group selection while both the group and lasso penalties have weaknesses. First, the group penalty is incapable of selecting specific covariate-taxon associations and is therefore guaranteed to have poor within-group precision, which remained at roughly 25\% for each scenario. On the other hand, while lasso is capable of selecting specific covariate-taxon associations, it suffered in terms of group selection due to worse group precision when compared with the other two penalties. This implies that the lasso penalty is more likely to keep covariates in the model that are truly irrelevant across the whole composition. In addition, the lasso penalty appears to be too conservative as it tended to have the lowest within-group recall of the three penalties.  As a combination of the group and lasso penalties, the sparse group lasso generally had higher within-group precision than the group penalty and higher group precision than the lasso penalty while maintaining higher within-group recall compared to the lasso penalty. Notably, DHR and MGLM demonstrated nearly identical performance, with only minuscule differences, when applied to DM regression with the same penalty function. Overall, these results demonstrate the importance of our DHR algorithm as it is capable of fitting regularized DM regression with SGL, unlike MGLM. SGL has proven to be more robust than group and lasso penalty functions, making it particularly beneficial for applications to regression models of compositional data.
\clearpage

\begingroup\fontsize{10}{12}\selectfont
\begin{longtable}[hp!]{rrllllllr}
\caption{\label{tab:table}Mean (SD) group and within-group selection performance of lasso, sparse group, and group penalties applied to Dirichlet-multinomial regression using the Dominating Hyperplane Regularization algorithm or the MGLM package in R, each with warm starts and convergence tolerance = $1e-5$. Performance is evaluated across simulation scenarios with varying numbers of candidate predictors ($p$) and proportions of relevant covariate associations ($\delta_p$), with $f=0.8$, $n=300$, $\delta_D=0.25$, and $D=12$. Results are averaged over 100 data replicates.}\\
\toprule
\multicolumn{4}{c}{ } & \multicolumn{2}{c}{Group Selection} & \multicolumn{2}{c}{Within-Group Selection} & \multicolumn{1}{c}{ } \\
\cmidrule(l{3pt}r{3pt}){5-6} \cmidrule(l{3pt}r{3pt}){7-8}
p & $\delta_p$ & Penalty & Method & Recall & Precision & Recall & Precision & Direction accuracy\\
\midrule
\endfirsthead
\caption[]{\textbf{(cont'd).}}\\
\toprule
\multicolumn{4}{c}{ } & \multicolumn{2}{c}{Group Selection} & \multicolumn{2}{c}{Within-Group Selection} & \multicolumn{1}{c}{ } \\
\cmidrule(l{3pt}r{3pt}){5-6} \cmidrule(l{3pt}r{3pt}){7-8}
p & $\delta_p$ & Penalty & Method & Recall & Precision & Recall & Precision & Direction accuracy\\
\midrule
\endhead

\endfoot
\bottomrule
\endlastfoot
 &  &  & DHR & 1.00 (0.00) & 0.93 (0.17) & 0.36 (0.09) & 0.96 (0.11) & 1.00\\

 &  & \multirow{-3}{*}{\raggedright\arraybackslash Lasso} & MGLM & 1.00 (0.00) & 0.91 (0.18) & 0.36 (0.09) & 0.98 (0.08) & 1.00\\

 &  &  & DHR & 1.00 (0.00) & 0.90 (0.19) & 0.52 (0.23) & 0.85 (0.17) & 1.00\\

 &  & \multirow{-3}{*}{\raggedright\arraybackslash SGL} & MGLM & NA (NA) & NA (NA) & NA (NA) & NA (NA) & NA\\

 &  &  & DHR & 1.00 (0.00) & 0.83 (0.18) & 1.00 (0.00) & 0.26 (0.01) & 1.00\\

 & \multirow{-10}{*}[1\dimexpr\aboverulesep+\belowrulesep+\cmidrulewidth]{\raggedleft\arraybackslash 0.10} & \multirow{-3}{*}{\raggedright\arraybackslash Group} & MGLM & 1.00 (0.00) & 0.99 (0.05) & 1.00 (0.00) & 0.26 (0.01) & 1.00\\
 \addlinespace

 &  &  & DHR & 1.00 (0.00) & 0.60 (0.18) & 0.88 (0.09) & 0.89 (0.06) & 1.00\\

 &  & \multirow{-3}{*}{\raggedright\arraybackslash Lasso} & MGLM & 1.00 (0.00) & 0.60 (0.18) & 0.87 (0.09) & 0.89 (0.06) & 1.00\\

 &  &  & DHR & 1.00 (0.00) & 0.69 (0.14) & 0.95 (0.06) & 0.81 (0.07) & 1.00\\

 &  & \multirow{-3}{*}{\raggedright\arraybackslash SGL} & MGLM & NA (NA) & NA (NA) & NA (NA) & NA (NA) & NA\\

 &  &  & DHR & 1.00 (0.00) & 0.92 (0.09) & 1.00 (0.00) & 0.25 (0.00) & 1.00\\

 & \multirow{-10}{*}[1\dimexpr\aboverulesep+\belowrulesep+\cmidrulewidth]{\raggedleft\arraybackslash 0.25} & \multirow{-3}{*}{\raggedright\arraybackslash Group} & MGLM & 1.00 (0.00) & 0.95 (0.07) & 1.00 (0.00) & 0.25 (0.00) & 1.00\\
 \addlinespace

 &  &  & DHR & 1.00 (0.00) & 0.60 (0.07) & 1.00 (0.01) & 0.80 (0.06) & 1.00\\

 &  & \multirow{-3}{*}{\raggedright\arraybackslash Lasso} & MGLM & 1.00 (0.00) & 0.60 (0.06) & 1.00 (0.01) & 0.80 (0.06) & 1.00\\

 &  &  & DHR & 1.00 (0.00) & 0.71 (0.08) & 1.00 (0.01) & 0.74 (0.05) & 1.00\\

 &  & \multirow{-3}{*}{\raggedright\arraybackslash SGL} & MGLM & NA (NA) & NA (NA) & NA (NA) & NA (NA) & NA\\

 &  &  & DHR & 1.00 (0.00) & 0.89 (0.07) & 1.00 (0.00) & 0.25 (0.00) & 1.00\\

\multirow{-33}{*}[4\dimexpr\aboverulesep+\belowrulesep+\cmidrulewidth]{\raggedleft\arraybackslash 25} & \multirow{-10}{*}[1\dimexpr\aboverulesep+\belowrulesep+\cmidrulewidth]{\raggedleft\arraybackslash 0.50} & \multirow{-3}{*}{\raggedright\arraybackslash Group} & MGLM & 1.00 (0.00) & 0.92 (0.06) & 1.00 (0.00) & 0.25 (0.00) & 1.00\\
\addlinespace
\hline
\addlinespace
 &  &  & DHR & 1.00 (0.00) & 0.62 (0.23) & 0.75 (0.18) & 0.97 (0.05) & 1.00\\

 &  & \multirow{-3}{*}{\raggedright\arraybackslash Lasso} & MGLM & 1.00 (0.00) & 0.64 (0.21) & 0.72 (0.17) & 0.97 (0.05) & 1.00\\

 &  &  & DHR & 1.00 (0.00) & 0.66 (0.16) & 0.96 (0.07) & 0.91 (0.07) & 1.00\\

 &  & \multirow{-3}{*}{\raggedright\arraybackslash SGL} & MGLM & NA (NA) & NA (NA) & NA (NA) & NA (NA) & NA\\

 &  &  & DHR & 1.00 (0.00) & 0.79 (0.12) & 1.00 (0.00) & 0.26 (0.00) & 1.00\\

 & \multirow{-10}{*}[1\dimexpr\aboverulesep+\belowrulesep+\cmidrulewidth]{\raggedleft\arraybackslash 0.10} & \multirow{-3}{*}{\raggedright\arraybackslash Group} & MGLM & 1.00 (0.00) & 0.95 (0.08) & 1.00 (0.00) & 0.26 (0.00) & 1.00\\
 \addlinespace

 &  &  & DHR & 1.00 (0.00) & 0.58 (0.12) & 0.87 (0.08) & 0.86 (0.05) & 1.00\\

 &  & \multirow{-3}{*}{\raggedright\arraybackslash Lasso} & MGLM & 1.00 (0.00) & 0.59 (0.13) & 0.85 (0.08) & 0.87 (0.05) & 1.00\\

 &  &  & DHR & 1.00 (0.00) & 0.69 (0.11) & 0.94 (0.05) & 0.79 (0.05) & 1.00\\

 &  & \multirow{-3}{*}{\raggedright\arraybackslash SGL} & MGLM & NA (NA) & NA (NA) & NA (NA) & NA (NA) & NA\\

 &  &  & DHR & 1.00 (0.00) & 0.91 (0.07) & 1.00 (0.00) & 0.25 (0.00) & 1.00\\

 & \multirow{-10}{*}[1\dimexpr\aboverulesep+\belowrulesep+\cmidrulewidth]{\raggedleft\arraybackslash 0.25} & \multirow{-3}{*}{\raggedright\arraybackslash Group} & MGLM & 1.00 (0.00) & 0.91 (0.06) & 1.00 (0.00) & 0.25 (0.00) & 1.00\\
 \addlinespace

 &  &  & DHR & 1.00 (0.00) & 0.61 (0.05) & 0.96 (0.03) & 0.78 (0.04) & 1.00\\

 &  & \multirow{-3}{*}{\raggedright\arraybackslash Lasso} & MGLM & 1.00 (0.00) & 0.61 (0.05) & 0.95 (0.04) & 0.78 (0.04) & 1.00\\

 &  &  & DHR & 1.00 (0.01) & 0.72 (0.06) & 0.97 (0.02) & 0.73 (0.04) & 1.00\\

 &  & \multirow{-3}{*}{\raggedright\arraybackslash SGL} & MGLM & NA (NA) & NA (NA) & NA (NA) & NA (NA) & NA\\

 &  &  & DHR & 1.00 (0.01) & 0.89 (0.05) & 1.00 (0.00) & 0.25 (0.00) & 1.00\\

\multirow{-33}{*}[4\dimexpr\aboverulesep+\belowrulesep+\cmidrulewidth]{\raggedleft\arraybackslash 50} & \multirow{-10}{*}[1\dimexpr\aboverulesep+\belowrulesep+\cmidrulewidth]{\raggedleft\arraybackslash 0.50} & \multirow{-3}{*}{\raggedright\arraybackslash Group} & MGLM & 1.00 (0.01) & 0.89 (0.05) & 1.00 (0.00) & 0.25 (0.00) & 1.00\\
\addlinespace
\hline
\addlinespace
\clearpage
 &  &  & DHR & 1.00 (0.00) & 0.68 (0.17) & 0.70 (0.11) & 0.94 (0.05) & 1.00\\

 &  & \multirow{-3}{*}{\raggedright\arraybackslash Lasso} & MGLM & 1.00 (0.00) & 0.71 (0.14) & 0.65 (0.08) & 0.94 (0.05) & 1.00\\

 &  &  & DHR & 1.00 (0.00) & 0.72 (0.14) & 0.89 (0.10) & 0.86 (0.06) & 1.00\\

 &  & \multirow{-3}{*}{\raggedright\arraybackslash SGL} & MGLM & NA (NA) & NA (NA) & NA (NA) & NA (NA) & NA\\

 &  &  & DHR & 1.00 (0.00) & 0.88 (0.10) & 1.00 (0.00) & 0.25 (0.00) & 1.00\\

 & \multirow{-10}{*}[1\dimexpr\aboverulesep+\belowrulesep+\cmidrulewidth]{\raggedleft\arraybackslash 0.10} & \multirow{-3}{*}{\raggedright\arraybackslash Group} & MGLM & 1.00 (0.00) & 0.93 (0.06) & 1.00 (0.00) & 0.25 (0.00) & 1.00\\
 \addlinespace

 &  &  & DHR & 1.00 (0.00) & 0.56 (0.08) & 0.81 (0.08) & 0.87 (0.04) & 0.99\\

 &  & \multirow{-3}{*}{\raggedright\arraybackslash Lasso} & MGLM & 1.00 (0.00) & 0.57 (0.08) & 0.79 (0.07) & 0.88 (0.04) & 1.00\\

 &  &  & DHR & 1.00 (0.00) & 0.65 (0.08) & 0.90 (0.05) & 0.79 (0.04) & 0.99\\

 &  & \multirow{-3}{*}{\raggedright\arraybackslash SGL} & MGLM & NA (NA) & NA (NA) & NA (NA) & NA (NA) & NA\\

 &  &  & DHR & 0.97 (0.18) & 0.91 (0.05) & 1.00 (0.00) & 0.25 (0.00) & 0.97\\

 & \multirow{-10}{*}[1\dimexpr\aboverulesep+\belowrulesep+\cmidrulewidth]{\raggedleft\arraybackslash 0.25} & \multirow{-3}{*}{\raggedright\arraybackslash Group} & MGLM & 1.00 (0.00) & 0.90 (0.05) & 1.00 (0.00) & 0.25 (0.00) & 0.99\\
 \addlinespace

 &  &  & DHR & 1.00 (0.00) & 0.60 (0.07) & 0.93 (0.06) & 0.74 (0.05) & 0.99\\

 &  & \multirow{-3}{*}{\raggedright\arraybackslash Lasso} & MGLM & 1.00 (0.00) & 0.57 (0.05) & 0.94 (0.06) & 0.72 (0.04) & 1.00\\

 &  &  & DHR & 1.00 (0.00) & 0.68 (0.08) & 0.95 (0.04) & 0.68 (0.04) & 0.99\\

 &  & \multirow{-3}{*}{\raggedright\arraybackslash SGL} & MGLM & NA (NA) & NA (NA) & NA (NA) & NA (NA) & NA\\

 &  &  & DHR & 0.00 (0.00) & NA (NA) & NA (NA) & NA (NA) & NA\\

\multirow{-33}{*}[4\dimexpr\aboverulesep+\belowrulesep+\cmidrulewidth]{\raggedleft\arraybackslash 100} & \multirow{-10}{*}[1\dimexpr\aboverulesep+\belowrulesep+\cmidrulewidth]{\raggedleft\arraybackslash 0.50} & \multirow{-3}{*}{\raggedright\arraybackslash Group} & MGLM & 0.22 (0.12) & 1.00 (0.00) & 0.98 (0.03) & 0.29 (0.02) & 0.81\\*

\end{longtable}
\endgroup{}

\end{document}